\documentclass[]{aastex631}

\usepackage{ulem}
\usepackage{soul}
\usepackage{amsmath}
\usepackage{makecell}
\usepackage{cancel}
\usepackage{color}

\shorttitle{Modeling COMs Formation in cold core: Multi-phase Models with Non-thermal Mechanisms}
\shortauthors{Yang Lu et al.}

\graphicspath{{./}{figures/}}

\begin{document}

\title{Modeling Complex Organic Molecules Formation in Cold Cores: Multi-phase Models with Non-thermal Mechanisms}

\correspondingauthor{Donghui Quan}
\email{donghui.quan@zhejianglab.com}

\author[0000-0002-0595-8918]{Yang Lu}
\affiliation{Research Center for Astronomical Computing, Zhejiang Laboratory, Hangzhou 311100, China}

\author[0000-0003-4811-2581]{Donghui Quan}
\affiliation{Research Center for Astronomical Computing, Zhejiang Laboratory, Hangzhou 311100, China}

\author[0000-0002-0040-6022]{Qiang Chang}
\affiliation{School of Physics and Optoelectronic Engineering, Shandong University of Technology, Zibo 255000, China}

\author[0000-0002-9703-3110]{Long-Fei Chen}
\affiliation{School of Physics and Electronic Science, Guizhou Normal University, Guiyang 550025, China}

\author[0000-0003-3010-7661]{Di Li}
\affiliation{Research Center for Astronomical Computing, Zhejiang Laboratory, Hangzhou 311100, China}
\affiliation{National Astronomical Observatories, Chinese Academy of Sciences, 20A Datun Road, Chaoyang District, Beijing 100101, China}
\affiliation{School of Astronomy and Space Science, University of Chinese Academy of Sciences, Beijing 100049, China}

\begin{abstract}
In recent years, a significant number of oxygen-bearing complex organic molecules (COMs) have been detected in the gas phase of cold dark clouds such as TMC-1. 
The formation of these COMs cannot be explained by diffusive mechanisms on grains and gas phase reactions.
This study investigates the formation of oxygen-bearing COMs in cold dark clouds using multiphase gas-grain models that incorporate cosmic ray-induced non-diffusive radiation chemistry and non-thermal sputtering desorption mechanisms. 
Additionally, we present the effects of varying elemental C/O ratio and different sputtering rates.
We utilized an accelerated Gillespie algorithm, based on the regular Gillespie algorithm.
The results of our models for dimethyl ether (CH$_{3}$OCH$_{3}$), methyl formate (HCOOCH$_3$), acetaldehyde (CH$_3$CHO), ethanol (C$_2$H$_5$OH), and methanol (CH$_3$OH) show reasonable agreement with observations toward TMC-1, within a factor of 3. Out of the 94 species compared with observations, 63 show agreement within 1 order of magnitude, accounting for 67.02\%. Overall inclusion of non-thermal mechanisms in multi-phase models shows notable improvement of modeling on oxygen-bearing COMs in the interstellar medium.
\end{abstract}


\section{Introduction} \label{sec:intro}
Complex organic molecules (COMs), which are defined as carbon-containing molecules with at least six atoms \citep{2009ARAA..47..427H}, are primarily observed in low-mass \citep{Bottinelli:2004ki,Jorgensen:2012fp,2020AA...639A..87V} and high-mass star-forming regions \citep{Belloche:2013rsa,doi:10.1126/science.1256678,2018ApJS..237....3S,2021ApJ...909..214L}.
The synthesis mechanism of COMs in the warm environment of the interstellar medium (ISM) is widely studied. Radicals produced through photodissociation on the dust surfaces and mantles undergo diffusion and recombination processes, leading to the formation of COMs at temperatures ranging from 20 to 40 K \citep{Garrod:2006zp,Garrod:2008wy,2014ApJ...787..135C}. Subsequently, as the temperature of the grains increases, the formed COMs are sublimated \citep{1992ApJ...399L..71C}. Additionally, gas phase reactions could also contribute to the synthesis of COMs \citep{2015MNRAS.449L..16B,2017ApJ...842...33V}.

In recent years, many oxygen-bearing COMs, such as CH$_{3}$OCH$_{3}$, HCOOCH$_3$, CH$_3$CHO and CH$_3$OH, have been detected in cold environments, including sources like L1689B \citep{2012AA...541L..12B,2016AA...587A.130B}, B1-b \citep{2012ApJ...759L..43C,2010ApJ...716..825O}, L1544 \citep{2016ApJ...830L...6J,2014ApJ...795L...2V}, L1498 \citep{2021ApJ...917...44J}, TMC-1 \citep{2021AA...649L...4A,2020AA...642L..17C}, L1989B \citep{2012AA...541L..12B,2015EAS....75..325B}, Barnard 5 \citep{2017AA...607A..20T}, and L483 \citep{2019AA...625A.147A}.
In cold environments, with temperatures as low as approximately 10 K, the diffusion of the majority of species is notably hindered, with only a few exceptions, such as hydrogen atoms.
The detection of COMs in these cold environments challenges current astrochemical models relying on diffusive grain surface and mantle chemistry.
While some studies suggest that diffusive grain surface chemistry may form certain COMs more efficiently at 10 K than previously thought \citep{2024AA...688A.150M,2024ApJ...974..115F},
other studies have indicated that these COMs were synthesized on dust through non-diffusive mechanisms and subsequently desorbed into the gas phase through non-thermal desorption mechanisms \citep{2021AA...652A..63W,2022MNRAS.516.4097P}.

Various non-diffusive mechanisms have been proposed to explain the observed COMs in cold cores.
These mechanisms include:
(1) Eley-Rideal Mechanism:
gas phase species directly react with grain surface species at the same site upon adsorption onto the grain, bypassing the need for surface diffusion \citep{2015MNRAS.447.4004R}.
(2) Chain Reaction Mechanism: a radical (O) reacts with diffusive radicals (H), followed by subsequent reactions with other radicals (CO) without requiring diffusion \citep{2016ApJ...819..145C}.
(3) Oxygen Insertion Reaction Mechanism: driven by cosmic rays or UV radiation, oxygen atoms are excited and undergo barrier-free insertion reactions with adjacent species, leading to the formation of oxygen-bearing products on grain surfaces or in the gas phase \citep{2017ApJ...845...29B,2023MNRAS.519.4622C}.
(4) Cosmic-Ray-Induced Radiolysis:
cosmic rays can induce molecules within grain surfaces and bulk ice into electronically excited states, enabling them to overcome reaction energy barriers and react with neighboring molecules \citep{2018ApJ...861...20S,2021MNRAS.500.3414P}.
(5) Gas phase formation through radical reactions: new pathways for synthesizing COMs in the gas phase \citep{2015MNRAS.449L..16B}.
(6) Three-body reaction mechanism: a chemical reaction product can react with a nearby molecule. \citep{2020ApJS..249...26J}.

At the same time, some non-thermal desorption mechanisms have also been studied, such as:
(1) Chemical desorption: the energy released during exothermic grain surface reactions can lead to the partial evaporation of the reaction products \citep{2007...467.1103G,2013NatSR...3E1338D,2016AA...585A..24M,2017MolAs...6...22W,2021AA...652A..63W}.
(2) UV-Induced Photodesorption: the absorption of a high-energy UV photon by species near a grain surface leads to the desorption of those species as their binding energy is exceeded \citep{2007ApJ...662L..23O,2016ApJ...817L..12B,2016AA...592A..68C}.
(3) Cosmic-Ray-Induced grain heating desorption: the collision of cosmic-ray particles with grains causes localized or global heating, leading to species desorption \citep{2004AA...415..203S}.
(4) Cosmic-Ray-Induced sputtering desorption: high-energy particles impacts from cosmic rays directly result in the ejection of species from the grain surface and bulk ice
\citep{2018AA...618A.173D,2019AA...627A..55D,2021AA...647A.177D,2022MNRAS.516.4097P,2023MNRAS.518.2050A}.

The aforementioned mechanisms were investigated within the framework of the three-phase model of the rate equation, such as Nautilus \citep{2016MNRAS.459.3756R}, which has shown promising results in enhancing the abundance of COMs in the gas phase. However, there were still notable discrepancies between the model simulations and observational data. For example,
\cite{2022MNRAS.516.4097P} introduced a theoretical sputtering rate evaluation method and applied it to a rate equation based model of cold dark clouds. The focus was on the impact of cosmic ray sputtering rate of grain ice surfaces composed of water, carbon dioxide, and a simple mixed ice on the abundance of gas phase COMs. They integrated radiolysis mechanisms into their simulations, allowing COMs to form on dust at low temperatures. The results indicated an effective enhancement in the abundance of gas phase COMs. However, among the seven models examined, there was an underproduction of HCOOCH$_3$ compared to observations, with levels lower by one to three orders of magnitude. Through a supplementary model that increased the sputtering rate, the abundance of HCOOCH$_3$ could be raised, but this adjustment led to an excess production of CH$_3$OCH$_3$, surpassing observed levels by approximately one order of magnitude. \cite{2020ApJS..249...26J} employed various nondiffusive mechanisms combined with chemical desorption mechanisms to investigate the formation of COMs in the prestellar core L1544. The results demonstrated a significant enhancement in the abundance of COMs in the gas phase relative to previous models. However, in the ``3-BEF Best'' model, while the simulated abundances of CH$_3$CHO and HCOOCH$_3$ closely matched observations, the abundance of CH$_3$OCH$_3$ was underestimated by two orders of magnitude. In the ``3-B+3-BEF" model, when the simulated abundance of CH$_3$CHO aligned with observations, the abundance of CH$_3$OCH$_3$ was underestimated by two orders of magnitude, whereas the abundance of HCOOCH$_3$ was overestimated by two orders of magnitude compared to observations. In our understanding, \cite{2021AA...652A..63W} was the first to apply experimentally measured sputtering rates of grain ice composed of CO$_2$ and H$_2$O to testing their efficiency in a rate equation based model of cold-core conditions. Sputtering is the primary desorption method at high densities for molecules that rely on dust formation, such as HCOOCH$_3$, CH$_3$OH. The introduction of non-thermal mechanisms alone resulted in simulation results of COMs in the gas phase significantly lower than the observed results, such as HCOOCH$_3$ and CH$_3$CHO.

In this study, we utilized the multiphase astrochemical model developed by \cite{2018ApJ...869..165L} and specifically incorporated radiation and sputtering reactions induced by cosmic rays. 
We applied experimentally measured different sputtering rates  \citep{2021AA...647A.177D}  to a multiphase astrochemical model of a cold cloud and tested its desorption efficiency for COMs within grain. By considering the sputtering rate of ice mantles composed of CO, CO$_2$ and H$_2$O. In  addition, a theoretical weighting method was also used to adjust the rate coefficient for sputtering based on the fractional amounts of CO, CO$_2$ and H$_2$O ice within the model. We have expanded the reaction network following radiation excitation. The excited molecules are not limited to synthesizing COMs, but have an equal probability of reacting with other potential molecules.
We utilized an accelerated Gillespie algorithm developed in our previous work  \citep{2017ApJ...851...68C}, to simulate the chemical evolution process of cold cores, enabling us to investigate the formation of oxygen-bearing COMs. 
We applied typical 0D cold molecular cloud physical conditions in our study. We compared the simulated results with observations of TMC-1, toward which approximately 150 different species were detected (see table \ref{obsTMC1} and \footnote{\url{https://cdms.astro.uni-koeln.de/classic/molecules}}). 
In addition, as a significant number of carbon chain species were observed in TMC-1, we also investigated the impact of different initial abundance values of C/O and compared the results with observed and previously studied, in order to study the effects of varying C/O ratios on the formation of carbon chain species and COMs.
The structure of this article is as follows, 
Section \ref{CHEMICAL MODELS} presents modeling methodologies, including the cosmic-ray-induced sputtering and radiolysis mechanisms employed in the multiphase model, along with the physical parameters and chemical parameters utilized. 
Section \ref{sec:Results} presents the results.
Section \ref{Conclusions1} provides a summary of our findings.

\section{Models and Mechanisms}
\label{CHEMICAL MODELS}
\subsection{Physical Parameters}
In this paper, we adopted a typical 0D cold cloud core physical model with constant physical parameters, as listed in table \ref{tab:physicalcond} \citep{2010AA...517A..21W,2013AA...550A..36M,2022MNRAS.516.4097P}. The density of hydrogen nuclei is $n_{\text{H}}= 2\times10^4$ $\text{cm}^{-3}$ \citep{2023AA...677A.106A}.
We used the standard dust size with a radius of r$_d$ = 0.1 $\mu$m and a surface density of sites of 1.5$\times 10^{15}$ sites per cm$^{2}$ \citep{1995ApJ...455..389J}. Therefore, each monolayer on the grain has approximately N$_s$ $\approx 10^6$ binding  sites. The gas phase and dust have the same temperature T$_{\rm{gas}}$ = T$_{\rm{dust}}$ = 10 K. The visual extinction A$_{\text{v}}$ = 10 mag.
We utilized a gas cell around a dust particle, as described by \citet{2009ApJ...691.1459V}. This gas cell contains a total of 10$^{12}$ H nuclei, with a dust grain density set at 3 g cm$^{-3}$, and the dust-to-gas mass ratio is maintained at 0.01.
The standard cosmic-ray ionization rate of H$_2$ is employed, $\zeta=1.3 \times 10^{-17}$ s$^{-1}$.
\begin{deluxetable*}{ll}
\tablecaption{Physical Parameters Utilized in Models (Based on TMC-1 Conditions).}
\tabletypesize{\scriptsize}
\label{tab:physicalcond}
\tablehead{\colhead{Parameter} &\colhead{TMC1} } 
\startdata
$\textit{n}_{\textup{H}}$ (cm$^{-3}$) & $2 \times10^{4}$ \\
r$_d$($\mu$m) &0.1\\
$\text{T}_{\textup{gas}}$ (K) & $10$ \\
$\text{T}_{\textup{dust}}$ (K) & $10$ \\
$\textit{N}_\textup{{site}}$ (cm$^{-2}$) & $1.5 \times 10^{15}$ \\
$\zeta$ (s$^{-1}$) & $1.3 \times 10^{-17}$ \\
$\rm{A}_{\text{v}}$(mag)&10
\enddata
\end{deluxetable*}
\subsection{Chemical Models}
\begin{figure}
	\includegraphics[width=\columnwidth]{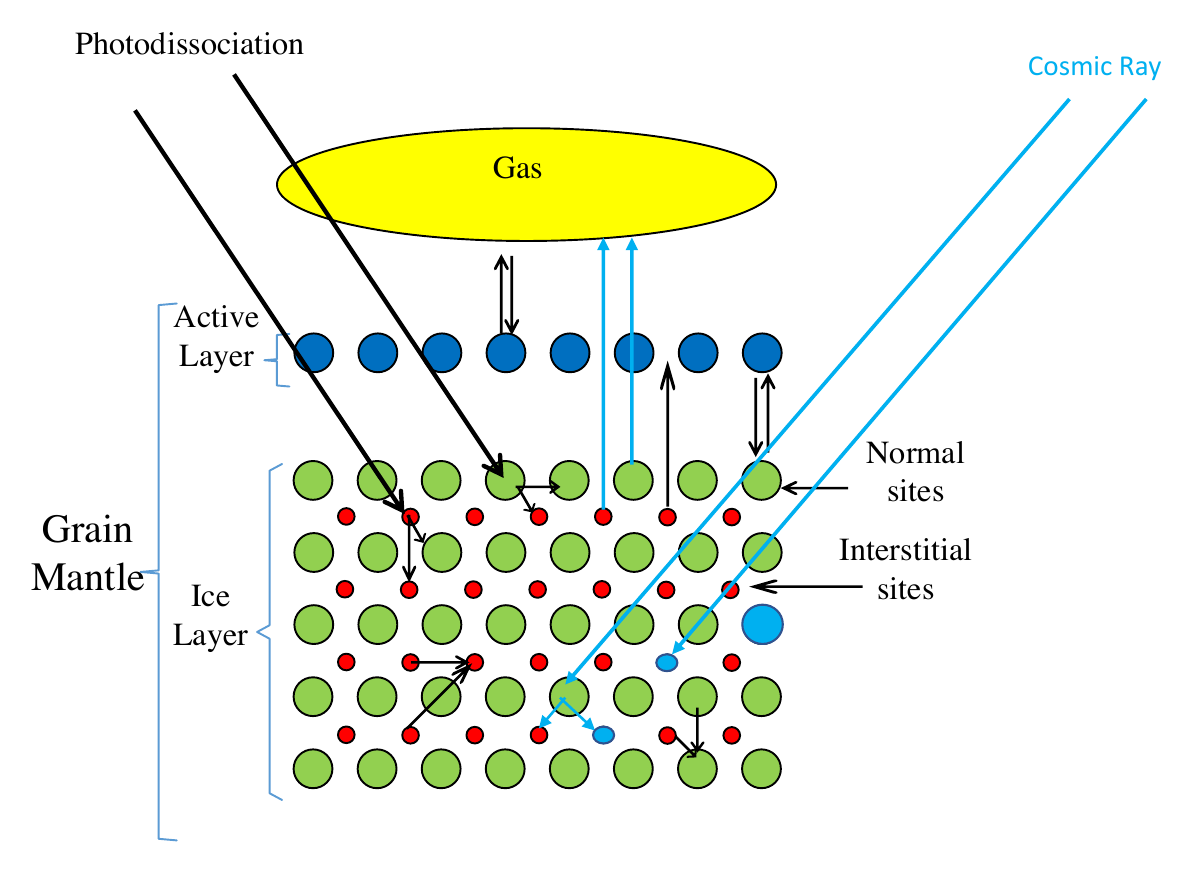}
    \caption{The schematic diagram depicts the multiphase model. The yellow area represents the gas phase. The blue circles represent the active layer molecules, capable of freely diffusing and undergoing reactions. The green circles depict the normal species within the frozen ice mantle, which are unable to diffuse. The red circles represent the interstitial species, capable of diffusion and reacting with both interstitial and normal species. The presence of light blue circles indicates that the molecules have been excited by cosmic rays, resulting in an excited state that enables them to react with neighboring molecules. Cosmic rays can cause molecules, including those in interstitial and normal species, to be sputtering into the gas phase.}
    \label{fig:model_schematic_diagram}
\end{figure}

The multiphase chemical model used in this paper was extensively described in our previous research \citep{2018ApJ...869..165L}.  
Here, we provide a brief overview of the model and introduce the newly introduced cosmic ray-induced non-thermal chemistry processes in the model. Figure \ref{fig:model_schematic_diagram} shows a schematic diagram of the multiphase model containing the new non-thermal mechanism.
The model comprises of the gas phase, dust surface, and dust bulk ice.
The dust surface, also known as the active layer, is composed of the topmost four monolayers of dust, which can freely diffuse and react with each other.
The active layer, located between the gas phase and bulk ice, facilitates direct species exchange with both of them.
The bulk ice mantle is composed of multiple monolayers of normal species and interstitial species. The normal species are time-marked and bound in position, meaning they cannot diffuse. On the other hand, the interstitial species are uniformly distributed, capable of diffusion, and can react with both themselves and the normal species.
The model parameters used in this article were as follows:
the surface diffusion barrier, denoted as E$_b$, is fixed at 50\% of the desorption energy, E$_D$. Additionally, the diffusion barrier for interstitial species, denoted as E$_{b2}$, is calculated as 0.7 times the desorption energy. The probability ratio of photodissociation products entering the interstitial layer compared to the normal layer is denoted by $\boldsymbol{\alpha} =$ 0.5.
The light blue process represents our newly introduced mechanisms. 
The light blue circles indicate that the radiolysis process produces species in excited states, which can react directly with neighboring species without diffusion. The light blue arrow pointing to the gas phase indicates that species on the dust surface and in the ice mantle desorption into the gas phase through the sputtering mechanism. The detailed process and calculation methodology are presented in the following section.
In this study, we employed an accelerated Gillespie algorithm (QSSA1), developed by \citet{2017ApJ...851...68C}, to simulate the gas-grain chemical reaction network described by \citet{2018ApJ...869..165L}.
This network consisted of 464 gas species, 197 dust species, and a total of 6370 reactions.
Building upon this network, we incorporated an additional 46 radiolysis reactions (refer to Table \ref{Radiolysisreaction}), 343 suprathermal reactions (see Appendix \ref{appendix A} Table \ref{Suprathermalreaction}), and 197 sputtering reactions. Additionally, we consider 25 molecules that can be excited. We used the initial low-metallicity abundances for gas phase species \citep{2010AA...522A..42S}, as displayed in Table ~\ref{tab:abundances}. 

This study employs a macroscopic Monte Carlo multiphase chemical model that balances precision in modeling surface and ice mantle chemistry with computational efficiency.
In this context, we used the accelerated Gillespie algorithm to model the accretion and desorption of H$_2$ processes that were previously ignored in all Monte Carlo simulations of gas-grain reaction networks due to computational cost \citep{2017ApJ...851...68C}.
Compared to rate equation three-phase models (e.g., \cite{2020ApJS..249...26J,2021MNRAS.500.3414P}), this model better models the non-uniformity of ice layers, enabling a more accurate description of non-diffusive reactions on dust ice mantles (section 2.3). 
Microscopic Monte Carlo models provide a more precise method for simulating the structure of dust ice mantles, making them suitable for studying complex ice layer structures. With this approach, the position of each particle on the dust, modeled as a planar lattice, can be tracked, allowing for the investigation of processes like diffusion and desorption based on the local environment of each adsorbate \citep{2012ApJ...759..147C,2014ApJ...787..135C}. However, their high computational demands make them challenging to apply to larger reaction networks and longer chemical evolution timescales. For example, in simulations of cold dense interstellar clouds, \cite{2012ApJ...759..147C} employed a surface chemical network with only 29 reactions, and \cite{2009AA...508..275C} similarly used a limited dust reaction network. In contrast, \cite{2014ApJ...787..135C} implemented a full gas-grain network with approximately 300 dust reactions but could only simulate the system's evolution for around $2 \times 10^5$ years due to high computational costs. Therefore, this method is more difficult to apply to chemistry occurring during the warm-up phase of stellar evolution. Our model (without sputtering and radiation excitation mechanisms) has already been successfully applied to simulate the chemical evolution of lukewarm corinos \citep{2019AA...622A.185W}, protostellar cores \citep{2018ApJ...869..165L}, and massive star-forming regions \citep{2021AA...648A..72W}. The simulations are completed within a few days and show good agreement with observations. 
However, in this paper, simulating the evolution of cold cores through $1 \times 10^6$ years takes less than one day.

\begin{deluxetable*}{ll}
\tablecaption{Initial abundances of elements with respect to total hydrogen nuclei}
\tabletypesize{\scriptsize}
\label{tab:abundances}
\tablehead{\colhead{Species} &\colhead{Abundance} 
} 
\startdata
 He       & $9.0 \times 10^{-2}$ \\
  e$^-$    & $1.2 \times 10^{-4}$ \\
  H$_2$    & $5.0 \times 10^{-1}$ \\
  C$^{+}$  & $1.2 \times 10^{-4}$\\
  N        & $7.6 \times 10^{-5}$\\
  O        & $2.6 \times 10^{-4}$\\
  S$^{+}$  & $8.0 \times 10^{-8}$\\
  Si$^{+}$ & $8.0 \times 10^{-9}$\\
  Na$^{+}$ & $2.0 \times 10^{-9}$\\
  Mg$^{+}$ & $7.0 \times 10^{-9}$\\
  Fe$^{+}$ & $3.0 \times 10^{-9}$\\
  P$^{+}$  & $2.0 \times 10^{-10}$\\
  Cl$^{+}$ & $1.0 \times 10^{-9}$\\
\enddata
\end{deluxetable*}
\subsection{Radiolysis Mechanism by Cosmic Rays}
Cosmic rays are high-energy particles, including charged particles like protons, electrons, and heavy ions~\citep{2013AARv..21...70B,2016ApJ...831...18C}. 
They travel through space at nearly the speed of light and play a significant role in various astrophysical processes and particle physics~\citep{2016ApJ...831...18C}.
The incorporation of radiolysis chemical processes induced by cosmic rays into astrochemistry was first introduced by \cite{2016PNAS..113.7727A}. Subsequent investigations by \cite{2018PCCP...20.5359S} further elucidated these processes, identifying four distinct decomposition pathways that occur during radiolysis when cosmic rays interact with molecules on dust. For example, for a molecular species A,

\begin{align}
A \leadsto A^+ + e^- \tag{R1} \label{p1},\\
A \leadsto A^+ + e^- \rightarrow A^* \rightarrow B^* + C^* \tag{R2} \label{p2},\\
A \leadsto A^* \rightarrow B + C \tag{R3} \label{p3},\\
A \leadsto A^* \tag{R4} \label{p4}.
\end{align}

In the above reactions, the symbol * denotes an electronically excited species, which is also referred to as a ``suprathermal" species.
Suprathermal species have an extremely short existence time ($\ll$1s) \citep{2016ApJ...831...18C,2018ApJ...852...70B}, much briefer than the average thermal diffusion timescale of species ($\gg$1s at 10K)~\citep{1992ApJS...82..167H}.
Consequently, in cold environments ($\sim$10 K), these suprathermal species either react with neighboring molecules or quickly return to the ground state.

We incorporated this mechanism into the multiphase model, where both the molecules on the dust surface and in the bulk ice could undergo these reactions.
The light blue circles in Figure~\ref{fig:model_schematic_diagram} represent species that were excited by cosmic rays and could react with neighboring species.
We defined the dust species Ji to include surface species i (gi), interstitial species i (Ii), and species i locked in the normal sites (Ki).
When a molecule undergoes a radiolysis reaction, its location within the ice layer needs to be determined. 
Due to limited knowledge of how ice thickness affects radiolysis, 
we assume that radiolysis can occur in any layer of the ice mantle,
without accounting for depth-dependent attenuation.
When a molecule $i$ undergoes a radiolysis reaction, the distribution probability of molecule $i$ in the $j$-th ice layer is calculated as $P_{ji}$ = $SUMK_{ji}$/$\sum_{j}$$SUMK_{ji}$, where $SUMK_{ji}$
represents the population of molecule $i$ in the $j$-th ice layer. A random number $r$ between 0 and 1 is generated to identify the location of the excited molecule. If $r < P_{ji}$, molecule $i$ is excited in the $j$-th ice layer. Next, we randomly select the position occupied by the molecule $i$ in the $j$-th ice layer, the location of this molecule $i$ becomes empty after the reaction. 
Empty normal sites are distributed non-uniformly. Allowing radiolysis products to randomly or proportionally occupy these sites would significantly increase computational demands. To simplify the simulation, we assumed that all radiolysis products enter interstitial sites. This approach streamlined the process and reduced computational costs, enabling each model to complete within one day on a 3.0 GHz core.
We start from the topmost layer of the ice and select downward to determine the location of the molecule.

If the radiative excitation reaction generates an excited molecule $i_{exc}$, it does not diffuse but instead reacts immediately with neighboring species. 
Such reactions, classified as non-diffusive reactions, are determined as follows, the probability $P$ of a species $x$ being located around $i_{exc}$ in the $j$-th interstitial ice layer is calculated as $P$ = $SUMI_{jx}$/$SUMI_{j,total}$, where $SUMI_{jx}$
represents the population of species $x$ in the $j$-th interstitial layer, and $SUMI_{j,total}$ is the total population of species in the $j$-th interstitial layer. A random number $r$ between 0 and 1 is generated. If $r < P$, the excited molecule $i_{exc}$ reacts with species 
$x$. If species $x$ cannot react with $i_{exc}$, the excited molecule $i_{exc}$ returns to its ground state.

Whenever radiolysis occurs, we dynamically update the species distribution within the ice layer. 
This dynamic updating of species distribution in the grain ice is similar to the photodissociation process described in our paper \citet{2018ApJ...869..165L}. 
We have neglected potential changes in the structural properties of the ice layer induced by radiolysis, such as alterations in the ice lattice and variations in the number of fixed binding sites per ice layer.

Table \ref{Radiolysisreaction} shows the radiolysis process of the 13 molecules and the corresponding rate coefficients used in this study. 
The table includes a total of 25 species capable of being excited.
\startlongtable
\clearpage
\begin{deluxetable*}{llllllll}
\label{Radiolysisreaction}
\tablecaption{Radiolysis Reactions}
\tabletypesize{\scriptsize}
\tablehead{
\colhead{reactant} & \colhead{}&\colhead{reactant} &\colhead{}&\colhead{product} &\colhead{} &
\colhead{product}  &\colhead{rate coefficient (s$^{-1}$)}
} 
\startdata
JH$_2$O &+ &CR & $\rightarrow$ &JO$^*$ &+ &JH$_2^*$ &5.3296$\times 10^{-16}$\\
JH$_2$O &+ &CR & $\rightarrow$ &JOH$^*$ &+ &JH$^*$&5.3296$\times 10^{-16}$\\
JH$_2$O &+ &CR & $\rightarrow$ &JOH &+ &JH&2.5137$\times 10^{-16}$\\
JH$_2$O &+ &CR & $\rightarrow$ &JH$_2$O$^*$&&&2.5137$\times 10^{-16}$\\
\hline
JO$_2$& +&CR & $\rightarrow$ &JO$^*$ &+&JO$^*$ &5.3296$\times 10^{-16}$\\
JO$_2$& +&CR & $\rightarrow$ &JO &+&JO&3.0763$\times 10^{-16}$\\
JO$_2$& +&CR & $\rightarrow$ &JO$_2^*$ &&&3.0763$\times 10^{-16}$\\
\hline
JO$_3$& +&CR & $\rightarrow$ &JO$_2^*$ &+&JO$^*$ &5.3296$\times 10^{-16}$\\
JO$_3$& +&CR & $\rightarrow$ &JO$_2$ &+&JO&5.8404$\times 10^{-16}$\\
JO$_3$& +&CR & $\rightarrow$ &JO$_3^*$ &&&5.8404$\times 10^{-16}$\\
\hline
JCO& +&CR & $\rightarrow$ &JC$^*$ &+&JO$^*$&5.3296$\times 10^{-16}$ \\
JCO& +&CR & $\rightarrow$ &JC &+&JO&1.8259$\times 10^{-16}$\\
JCO& +&CR & $\rightarrow$ &JCO$^*$ &&&1.8259$\times 10^{-16}$\\
\hline
JCO$_2$& +&CR & $\rightarrow$ &JCO$^*$ &+&JO$^*$&5.3296$\times 10^{-16}$ \\
JCO$_2$& +&CR & $\rightarrow$ &JCO &+&JO&1.7971$\times 10^{-16}$\\
JCO$_2$& +&CR & $\rightarrow$ &JCO$_2^*$ &&&1.7971$\times 10^{-16}$\\
\hline
JNO& +&CR & $\rightarrow$ &JN$^*$ &+&JO$^*$& 5.3296$\times 10^{-16}$\\
JNO& +&CR & $\rightarrow$ &JN &+&JO&2.7655$\times 10^{-16}$\\
JNO& +&CR & $\rightarrow$ &JNO$^*$ &&&2.7655$\times 10^{-16}$\\
\hline
JNO$_2$& +&CR & $\rightarrow$ &JNO$^*$ &+&JO$^*$& 5.3296$\times 10^{-16}$\\
JNO$_2$& +&CR & $\rightarrow$ &JNO &+&JO&1.7367$\times 10^{-16}$\\
JNO$_2$& +&CR & $\rightarrow$ &JNO$_2^*$ &&&1.7367$\times 10^{-16}$\\
\hline
JO$_2$H& +&CR & $\rightarrow$ &JOH$^*$ &+&JO$^*$&5.3296$\times 10^{-16}$ \\
JO$_2$H& +&CR & $\rightarrow$ &JOH &+&JO&5.3439$\times 10^{-16}$\\
JO$_2$H& +&CR & $\rightarrow$ &JOH$_2^*$ &&&5.3439$\times 10^{-16}$\\
\hline
JH$_2$O$_2$& +&CR & $\rightarrow$ &JOH$^*$ &+&JOH$^*$&5.3296$\times 10^{-16}$ \\
JH$_2$O$_2$& +&CR & $\rightarrow$ &JH$_2$O$^*$ &+&JO$^*$& 5.3296$\times 10^{-16}$\\
JH$_2$O$_2$& +&CR & $\rightarrow$ &JOH$^*$ &+&JOH&3.3036$\times 10^{-16}$\\
\hline
JNH$_3$& +&CR & $\rightarrow$ &JH$^*$ &+&JNH$_2^*$&5.3296$\times 10^{-16}$ \\
JNH$_3$& +&CR & $\rightarrow$ &JH$_2^*$ &+&JNH$^*$&5.3296$\times 10^{-16}$\\
JNH$_3$& +&CR & $\rightarrow$ &JH &+&JNH$_2$&3.9152$\times 10^{-16}$\\
JNH$_3$& +&CR & $\rightarrow$ &JNH$_3^*$ &&&3.9152$\times 10^{-16}$\\
\hline
JCH$_4$& +&CR & $\rightarrow$ &JH$^*$ &+&JCH$_3^*$&5.3296$\times 10^{-16}$ \\
JCH$_4$& +&CR & $\rightarrow$ &JH$_2^*$ &+&JCH$_2^*$&5.3296$\times 10^{-16}$\\
JCH$_4$& +&CR & $\rightarrow$ &JH &+&JCH$_3$&2.1655$\times 10^{-16}$\\
JCH$_4$& +&CR & $\rightarrow$ &JCH$_4^*$ &&&2.1655$\times 10^{-16}$\\
\hline
JH$_2$CO& +&CR & $\rightarrow$ &JH$^*$ &+&JHCO$^*$ &5.3296$\times 10^{-16}$\\
JH$_2$CO& +&CR & $\rightarrow$ &JH&+&JHCO&4.1871$\times 10^{-16}$\\
JH$_2$CO& +&CR & $\rightarrow$ &JH$_2$CO$^*$ &&&4.1871$\times 10^{-16}$\\
\hline
JCH$_3$OH& +&CR & $\rightarrow$ &JH$^*$ &+&JCH$_3$O$^*$&5.3296$\times 10^{-16}$ \\
JCH$_3$OH& +&CR & $\rightarrow$ &JCH$_2$OH$^*$ &+&JH$^*$&5.3296$\times 10^{-16}$\\
JCH$_3$OH& +&CR & $\rightarrow$ &JOH$^*$ &+&JCH$_3^*$&5.3296$\times 10^{-16}$\\
JCH$_3$OH& +&CR & $\rightarrow$ &JH&+&JCH$_3$O&2.2605$\times 10^{-16}$ \\
JCH$_3$OH& +&CR & $\rightarrow$ &JH&+&JCH$_2$OH&2.2605$\times 10^{-16}$\\
JCH$_3$OH& +&CR & $\rightarrow$ &JOH &+&JCH$_3$&2.2605$\times 10^{-16}$\\
JCH$_3$OH& +&CR & $\rightarrow$ &JCH$_3$OH$^*$ &&&2.2605$\times 10^{-16}$\\
\enddata
\tablecomments{From \citet{2018ApJ...861...20S}. A total of 25 molecules can be excited, including JO$^*$,JO$_3^*$, JH$_2^*$, JOH$^*$, JH$^*$, JH$_2$O$^*$, JO$_2^*$, JC$^*$, JCO$^*$, JCO$_2^*$, JN$^*$, JNO$^*$, JNO$_2^*$, JO$_2$H$^*$, JNH$^*$, JNH$_2^*$, JNH$_3^*$, JCH$_2^*$, JCH$_3^*$, JCH$_4^*$, JHCO$^*$, JH$_2$CO$^*$, JCH$_3$O$^*$, JCH$_2$OH$^*$, JCH$_3$OH$^*$. \\CR: cosmic rays.}
\end{deluxetable*}
The approach introduced by \citet{2018PCCP...20.5359S} was employed to assess the rate coefficients for radiolysis pathways in this study.
In our models, COMs were predominantly synthesized through reactions involving suprathermal species on the ice mantle of dust particles. For example, the major formation channels of methyl formate in this work are as follows:

\begin{eqnarray}
\rm JHCO^*+ JCH_3O  \rightarrow JHCOOCH_3,\\
\rm JHCO + JCH_3O^*  \rightarrow JHCOOCH_3.
\end{eqnarray}

In Table \ref{tab:mainchannels}, we present a list of the primary channels through which COMs are formed via the radiation excitation mechanism discussed in this paper. 
In previous studies, electronically excited molecules were primarily used to form COMs. For example, in the work by \cite{2021MNRAS.500.3414P}, CH$_3^*$ only reacted with HOCO to form CH$_3$COOH. However, in our study, CH$_3^*$ is involved in 20 different reactions. It not only contributes to the formation of COMs but also participates in the synthesis of the methylpolyyne family and simple molecules. 
For additional details on the excited reactions on dust used in this work, please refer to Appendix \ref{appendix A}.

\begin{deluxetable*}{ll}
\tablecaption{The major Formation Channels for COMs.}
\tabletypesize{\scriptsize}
\label{tab:mainchannels}
\tablehead{\colhead{Specie} &\colhead{Reaction}}
\startdata
JCH$_2$OHCHO  &  JCH$_2$OH$^*$    +  JHCO\\
   			     &  JCH$_2$OH           +  JHCO$^*$\\
 \hline
 JHCOOCH$_3$ & JCH$_3$O$^*$  + JHCO\\
                         &  JCH$_3$O + JHCO $^*$ \\
\hline
JCH$_3$OCH$_3$ &JCH$_3^*$       + JCH$_3$O\\
			&JCH$_3$       + JCH$_3$O$^*$\\
\hline
JC$_2$H$_5$OH  & JCH$_3^*$       + JCH$_2$OH\\
 			     & JCH$_3$       + JCH$_2$OH$^*$\\
\hline			
JCH$_3$CHO &	JCH$_3^*$       + JHCO\\		
			&JCH$_3$       + JHCO$^*$\\
\hline
JCH3OH  &  JCH$_2$OH$^*$/JCH$_3$O$^*$ +  JH  \\
		&  JCH$_2$OH/JCH$_3$O  +  JH$^*$  \\
\enddata
\end{deluxetable*}

Carbon also plays a crucial role in interstellar chemistry \citep{2023arXiv230315769T}. 
Cold molecular clouds have been observed to contain various carbon-chain species, encompassing simple linear chains, cyclic structures, and polycyclic aromatic hydrocarbons (PAHs) \citep{2024AA...682A.109M}. These relatively simple and small carbon-chain molecules are predominantly formed in the gas phase through ion-molecule and neutral-neutral reactions \citep{2023arXiv230315769T}.
Consequently, we also investigated the potential impact of the radiolysis mechanism on the formation of the hydrocarbon family (C$_n$H, n=1,2,3...), cyanopolyynes family (HC$_n$N, where n=1,3,5,7,…), and methylcyanopolyynes family (CH$_3$C$_n$N, where n=1,3,5,...).

In our model, the hydrocarbon family primarily forms in the gas phase through two reaction channels, as previously confirmed in other studies \citep{2013ChRv..113.8710A,2023arXiv230315769T,2023ApJ...944L..45R}.

\begin{eqnarray}
\rm C_{n-1}H_2 + C  \rightarrow C_nH + H, \label{eq:carb1}\\
\rm C_nH_2^{+} + e^{-}  \rightarrow  C_nH + H. \label{eq:carb2}
\end{eqnarray}

The cyanopolyynes family primarily forms through the following three main mechanisms:
\citep{1998AA...329.1156T,2016ApJ...817..147T,2018MNRAS.474.5068B},

Mechanism 1, recombination reaction,
\begin{eqnarray}
\rm H_2C_nN^+ + e^-  \rightarrow HC_nN + H. \label{eq:cyanopolyynef1_1}
\end{eqnarray}

Mechanism 2, CN radical reacts with hydrocarbon molecules,
\begin{eqnarray}
\rm C_{n-1}H_2 + CN  \rightarrow HC_{n}N + H. \label{eq:cyanopolyynef2_1}
\end{eqnarray}

Mechanism 3, the next-smallest cyanopolyyne reacts with hydrocarbon molecules,
\begin{eqnarray}
\rm HC_{n-2}N + C_2H  \rightarrow HC_{n}N + H. \label{eq:cyanopolyynef3_1}
\end{eqnarray}
In our model, the methylcyanopolyynes family was primarily synthesized in the gas phase through recombination reactions.
\begin{eqnarray}
\rm CH_{3}C_nNH^+ + e^-  \rightarrow CH_3C_nN + H. \label{eq:methylcyanopolyynes_1}
\end{eqnarray}

\subsection{Sputtering Mechanism by Cosmic Rays}
Sputtering has been experimentally verified \citep{2018AA...618A.173D,2020AA...634A.103D,2021AA...647A.177D,2023AA...671A.156D} and theoretically supported \citep{2022MNRAS.516.4097P,2021AA...652A..63W,2023MNRAS.518.2050A}  as an effective non-thermal mechanism for desorbing species from dust into the gas phase.
The rate coefficients of sputtering reactions are calculated based on experimental fitting formulas, as follows:
\begin{equation}
    k_{\text{scr}}=\frac{\zeta}{3\times10^{-17}} \times Y_{\text{eff}} \times \pi \times r_d^2 /N_s.
	\label{eq:sputteringrate}
\end{equation}
where
\begin{equation}
    Y_{\text{eff}} = \alpha \times (1-e^{-(\frac{n_{\text{layers}}}{\beta})^{\gamma}}).
	\label{eq:eff}
\end{equation}
Here, $Y_{\text{eff}}$  is the effective sputtering rate, which depends on the number of layers of dust ice ($n_{\text{layers}}$, \citet{2021AA...647A.177D}).
The symbol $\alpha$ represents the sputtering yield for ice, while $\beta$ and $\gamma$ are two parameters associated with the nature of the ice, both defined by \cite{2018AA...618A.173D,2021AA...647A.177D}.
The sputtering rate is subject to multiple uncertain factors, including the composition of the dust ice which significantly affected the stopping power of the target ice, thus plays a crucial role in determining the sputtering cross-section \citep{2015NIMPB.365..472D,2015NIMPB.365..477M}. Additionally, the composition and flux of cosmic rays also play a role, contributing to the overall uncertainty in the sputtering rate \citep{2023MNRAS.518.2050A}.
We simulated six different models to account for various sputtering rates. Detailed parameters for each model are provided in Table \ref{tab:model}. 

\begin{deluxetable*}{lllllll}
\tablecaption{Model with listings of different sputtering rate parameters and indications of whether cosmic ray radiolysis was included.}
\tabletypesize{\scriptsize}
\label{tab:model}
\tablehead{\colhead{Model} &\colhead{$\alpha$} & \colhead{$\beta$}& \colhead{$\gamma$}&
\colhead{Radiolysis}&\colhead{Sputtering}&\colhead{Ice mantles}
}
\startdata
      MC1 & - & -&-&On& On&Mixed ice\\
      MC2 & 40.1$^a$ & 75.8$^a$&0.69$^a$&On& On&CO\\
      MC3 & 21.9$^a$ & 56.3$^a$&0.60$^a$&On& On&CO$_2$\\
      MC4 & 3.63$^a$ & 3.25$^a$&0.57$^a$&On& On&H$_2$O\\
      MC5 & - & -&-&On& Off&-\\
      MC6 & - & -&-&Off& Off&-\\
\enddata
\tablecomments{\\
$^a$ \cite{2021AA...647A.177D}.\\}
\end{deluxetable*}

We defined Model 1 (MC1) as a multiphase model that incorporated radiolysis and sputtering mechanisms. Due to the absence of experiments measuring the sputtering rate of multi-component dust ice layers, in this model, the sputtering rate was determined through a weighted calculation. This involved multiplying the individual sputtering rates of the single component H$_2$O, CO, and CO$_2$ ice by their respective fractional amounts within the model. 
We used the following formula to calculate the sputtering rate coefficient for any species in dust,
\begin{equation}
    k_{\text{scr}}^{\text{mix}} =k_{\text{scr(CO)}} \frac{\text{N}_{\text{CO}}}{\text{N}_{\text{CO}}+\text{N}_{\text{CO}_2}+\text{N}_{\text{H}_2\text{O}}} 
   +k_{\text{scr(CO}_2)} \frac{\text{N}_{\text{CO}_2}}{\text{N}_{\text{CO}}+\text{N}_{\text{CO}_2}+\text{N}_{\text{H}_2\text{O}}}
      +k_{\text{scr(H}_2\text{O})} \frac{\text{N}_{\text{H}_2\text{O}}}{\text{N}_{\text{CO}}+\text{N}_{\text{CO}_2}+\text{N}_{\text{H}_2\text{O}}}.
        \label{eq:mixrate}
\end{equation}
here, N$_i$ represents the count of species $i$ in the ice, calculated in a real-time manner. $k_{\rm scr(i)}$ represents the theoretically fitted experimental sputtering rate for a single ice composition i, which could be computed using formula \ref{eq:sputteringrate}.
Model 2 (MC2), Model 3 (MC3), and Model 4 (MC4) all incorporated radiolysis and sputtering mechanisms. In Model 2, the sputtering rate of dust species was determined by fitting to experimental data obtained from a single ice component, CO. In Model 3, a single ice component composed of CO$_2$ was utilized, and in Model 4, a single ice component composed of H$_2$O was employed. The specific parameters can be found in Table \ref{tab:model}.
Model 5 (MC5) contains the radiolysis mechanism but lacks a sputtering mechanism, and both mechanisms were absent in Model 6 (MC6); in other words, it represents the original model described in the \citet{2018ApJ...869..165L} article.
All six models mentioned above incorporate non-thermal desorption mechanisms, including UV-induced photodesorption (with a yield assumed to be $1 \times 10^{-4}$ molecules per photons for all species \citep{2008AA...491..907A}) and cosmic-ray-induced grain heating desorption.

\section{Results and Discussions}
\label{sec:Results} 
\subsection{O-bearing COMs}
\begin{figure}
	\includegraphics[width=\columnwidth]{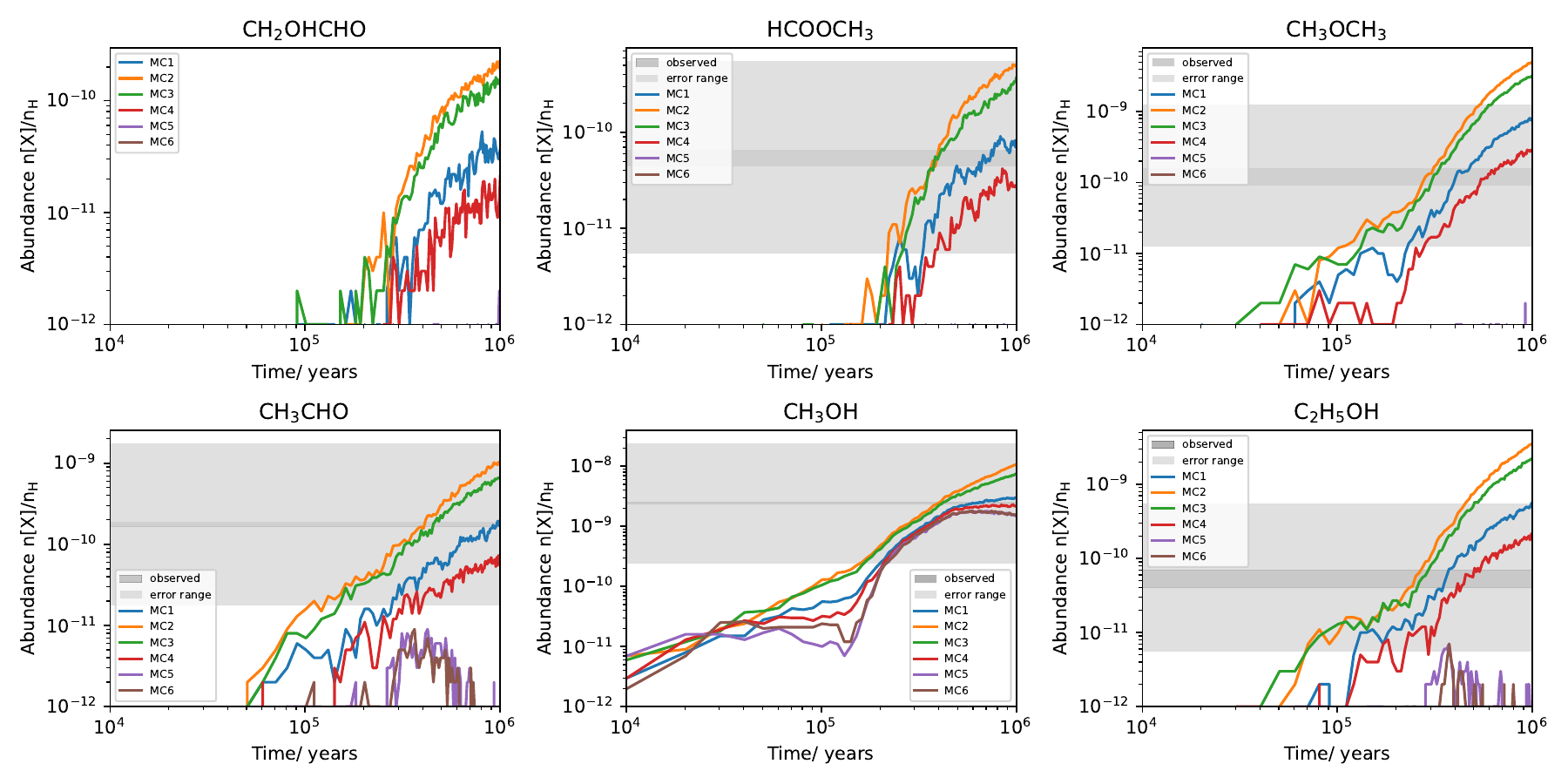}
    \caption{Abundance Ratios of Gas Species Relative to H in Different Models.}
    \label{fig:gasspec}
\end{figure}

Figure \ref{fig:gasspec} shows the temporal variation of selected gas phase oxygen-bearing COMs in models MC1, MC2, MC3, MC4, MC5, and MC6, alongside the observed results. The blue, yellow, green, red, purple, and brown solid lines represent the simulation results of MC1, MC2, MC3, MC4, MC5, and MC6 models, respectively. The dark gray area corresponds to the observed values (refer to Table \ref{tab:obesfit} for specific values), while the light gray area spans one order of magnitude above and below the observed values.

Our models, MC1 to MC4, which included new mechanisms, efficiently synthesized oxygen-bearing COMs in the gas phase, including CH$_2$OHCHO, HCOOCH$_3$, CH$_3$OCH$_3$, CH$_3$CHO, CH$_3$OH, and C$_2$H$_5$OH. The abundance of these COMs increased over time. The MC5 model, which contains only the radiolysis mechanism, and the MC6 model, which lacks both mechanisms, were only able to efficiently synthesize CH$_3$OH.
\subsubsection{CH$_3$OH}
All models are capable of forming CH$_3$OH in the gas phase. The primary formation pathway for gas phase CH$_3$OH is the hydrogenation of CO on grain surfaces to form JCH$_3$OH (gCO $\stackrel{\text{gH}} {\longrightarrow}$ gHCO  $\stackrel{\text{gH}} {\longrightarrow}$ gH$_2$CO  $\stackrel{\text{gH}} {\longrightarrow}$ gCH$_2$OH/gCH$_3$O  $\stackrel{\text{gH}} {\longrightarrow}$ JCH$_3$OH), which is then desorbed into the gas phase through non-thermal desorption mechanisms. The newly introduced radiolysis mechanism has a minimal impact on the formation of JCH$_3$OH, accounting for only about 1\% when the molecular cloud evolves to $1 \times 10^6$ years. However, the newly introduced sputtering mechanism alters the abundance distribution of species related to CH$_3$OH formation in both dust and gas phases, thereby increasing the efficiency of CH$_3$OH formation through the gas phase channel, with a maximum increase of approximately 11.98\% (MC2 model). Additionally, the sputtering desorption mechanism can desorb JCH$_3$OH into the gas phase from within the ice mantle and surface (whereas the photodesorption mechanism can only desorb CH$_3$OH from the dust surface), thereby enhancing the overall desorption efficiency of JCH$_3$OH and increasing its abundance in the gas phase. As shown in Figure \ref{fig:gasspec}, the abundance of CH$_3$OH in models MC1-MC4 gradually increases over time. Since MC1 and MC4 have lower sputtering rates compared to MC2 and MC3, the abundance of CH$_3$OH is lower in MC1 and MC4. They approach a steady state around $1 \times 10^6$ years, whereas MC2 and MC3 show a continuous gradual increase over time. At $1 \times 10^5$ years, the abundance of CH$_3$OH in models MC5 and MC6 declines due to ion-induced dissociation. In contrast, models incorporating sputtering mechanisms do not exhibit this decline because their total desorption rate surpasses the destruction rate. Eventually, the CH$_3$OH abundance in the MC5 and MC6 models reaches a steady state, remaining lower than that in the other four models.

In the MC1 model, at $1 \times 10^6$ years, approximately 6.18\% of CH$_3$OH is produced through gas phase reaction channels, with the main pathway being the dissociative recombination of CH$_3$OCH$_4^+$ (CH$_3$OCH$_4^+$ + $e^- \rightarrow $ CH$_3$OH $+$ CH$_3$). CH$_3$OCH$_4^+$ is mainly formed by the destruction of gas phase CH$_3$OCH$_3$ by ions (e.g., H$_3^+$, HCO$^+$, H$_3$O$^+$), with CH$_3$OCH$_3$ relying on the new mechanism for its formation on dust. 
CH$_3$OH desorbs from dust into the gas phase primarily through photodesorption, with sputtering acting as a secondary mechanism.
Cosmic-ray-induced grain heating desorption and thermal desorption are ineffective in the cold molecular cloud environment. 
The main formation pathway for JCH$_3$OH is the successive hydrogenation of CO on the dust surface, with a negligible contribution from radiolysis mechanisms.

In the MC2 model, the sputtering desorption rate is the highest, allowing more species from within the dust ice mantle to be desorbed into the gas phase. This significantly changes the abundance distribution of species in both the dust and gas phases and leads to an increase in the gas phase formation channel contribution to CH$_3$OH, which rises to approximately 11.98\%. More species, such as HCOOCH$_3$ and CH$_3$OCH$_3$, are desorbed into the gas phase, indirectly promoting the gas phase formation of CH$_3$OH (HCOOCH$_3$ $\rightarrow $ H$_5$C$_2$O$_2^+$ $\rightarrow $ CH$_3$OH, CH$_3$OCH$_3$ $\rightarrow $ CH$_3$OCH$_4^+$ $\rightarrow $ CH$_3$OH). 
In contrast to MC1, the MC2 model show that CH$_3$OH desorbs from dust into the gas phase primarily through sputtering, with photodesorption acting as a secondary mechanism. Most of the sputtering contribution originates from within the ice mantle.
Compared to other models, the MC2 model shows the highest abundance of CH$_3$OH in the gas phase.

In the MC3 model, the sputtering rate is second only to that of MC2. The abundance with temporal evolution of CH$_3$OH are similar to those in MC2, though slightly lower. 
Sputtering desorption is the dominant contributor to CH$_3$OH formation in the MC3 model, with photodesorption playing a secondary role. Gas-phase reactions contribute minimally, at approximately 10.49\%.

The MC4 model, with the lowest sputtering rate, also has the lowest CH$_3$OH abundance among the four models. Its evolution trend and abundance are similar to those of MC1, though it is about 8 times lower than in MC2. 
Similar to MC1, photodesorption is the primary process for CH$_3$OH formation in the MC4 model, with sputtering desorption serving as a secondary mechanism and gas-phase reactions contributing minimally.

In model MC5, which lack sputtering mechanism, and MC6, which lack both sputtering and radiolysis mechanisms, the CH$_3$OH  abundance profiles are nearly identical, particularly after $2 \times 10^5$ years, where their abundance curves overlap. In both models, CH$_3$OH  formed via photodesorption accounts for 98.41\% and 98.38\%, respectively. Their abundance is approximately an order of magnitude lower than in the MC2 model.

In all models, JCH$_3$OH is mainly formed on the dust surface through CO hydrogenation, with this process contributing to over 98\% of JCH$_3$OH formation. Radiolysis mechanisms have a negligible impact on its formation, accounting for only about 1\%. Consequently, even in the absence of radiolysis mechanisms in the MC6 model, the abundance of JCH$_3$OH  remains nearly consistent with that in the other five models. As shown in Figure \ref{fig:sputtering} the abundance of JCH$_3$OH in dust is consistent across all six models, with their abundance curves overlapping. 
JCH$_3$OH mainly forms on the dust surface, where its surface abundance is significantly higher than other surface COMs, by 3 to 4 orders of magnitude, 
close to the abundance of water on the surface of dust. Consequently, photodesorption of surface gCH$_3$OH alone can achieve a high gas-phase abundance, without relying on the sputtering of ICH$_3$OH or KCH$_3$OH from within the dust ice mantle. Therefore, models without sputtering and radiolysis mechanisms can still effectively synthesize CH$_3$OH, matching well with observed abundances. The introduction of new mechanisms has a minimal impact on the overall abundance of CH$_3$OH, increasing it by only one order of magnitude at the final time of 1$\times10^6$ years, while still aligning with observational data.
\subsubsection{HCOOCH$_3$}
HCOOCH$_3$ primarily forms within the dust ice mantle through radiative excitation mechanisms and is then desorbed into the gas phase via sputtering mechanism. 
In the MC1 to MC4 models, approximately 99\% of JHCOOCH$_3$ is formed on dust through radiative mechanisms, while around 99\% of the HCOOCH$_3$ in the gas phase is produced via sputtering mechanisms.
The reaction between JCH$_3$O and JHCO is the main formation pathway for JHCOOCH$_3$, but these molecules are rapidly hydrogenated on the grain surface, ultimately forming JCH$_3$OH. Therefore, HCOOCH$_3$ cannot effectively form on the dust surface. 
In our multiphase model, far-ultraviolet (FUV) photons can penetrate the ice layer, photodissociate species within the mantle, and produce free radicals (e.g., HCO, CH$_3$O). These free radicals can be frozen within the dust ice mantle, allowing them to persist for extended periods, while on the surface, these radicals are rapidly hydrogenated.
When molecules within the ice layer undergo radiolysis and enter an excited state, ICH$_3$O$^*$ (or IHCO$^*$) has a probability of reacting with IHCO (or ICH$_3$O) to form IHCOOCH$_3$. Unlike CH$_3$OH, the ice mantle is the primary formation site for HCOOCH$_3$.

Therefore, only the MC1 to MC4 models, which include both radiative and sputtering mechanisms, can effectively form HCOOCH$_3$ in the gas phase, as shown in Figure \ref{fig:gasspec}. 
In the MC2 model, the abundance of HCOOCH$_3$ is the highest among the four models, as it has the highest sputtering rate, allowing more HCOOCH$_3$ to desorb from the ice mantle into the gas phase. This is followed by the MC3, MC1, and MC4 models. In our model, the ice mantle begins to gradually form around $10^4$ years, reaching approximately 60 layers by 10$^5$ years (Figure \ref{fig:sputtering}). This allows a significant amount of HCO and CH$_3$O molecules to be frozen within the ice, enhancing their abundance in the mantle and then formation of HCOOCH$_3$. Consequently, HCOOCH$_3$ predominantly forms after 10$^5$ years.
When the simulated molecular cloud evolves to 1$\times 10^6$ years,
in the MC1 model, photodesorption contributes to about 0.5\% of HCOOCH$_3$ formation. In the MC2 model, this contribution is less than 0.1\%, while in the MC3 model, it is approximately 0.1\%. In the MC4 model, the contribution increases to around 1\%. Before 10$^5$ years, a small amount of HCOOCH$_3$ desorbs from the dust into the gas phase, but it is quickly destroyed by ions such as H$_3^+$, HCO$^+$, and H$_3$O$^+$, preventing its effective accumulation. 
When the simulated molecular cloud evolves to 1$\times 10^6$ years,
the recombination rate of HCOOCH$_3$ in the MC1 to MC4 models is approximately 50\% (HCOOCH$_3$ $\stackrel{\text{H$_3^+$,H$_3$O$^+$, HCO$^+$}}  \longrightarrow$ H$_5$C$_2$O$_2^+$ $\stackrel{\text{e$^-$}}  \longrightarrow$ HCOOCH$_3$).
The MC5 model, which includes the radiative mechanism, can effectively form JHCOOCH$_3$ on the grain, with its abundance showing a consistent trend with the other four models, peaking at approximately $1 \times 10^{-7}$, as shown in Figure \ref{fig:icespec}. However, due to the lack of sputtering mechanisms, the HCOOCH$_3$ formed on the grain ice mantle in the MC5 model cannot be desorbed into the gas phase. The MC6 model, which lacks both mechanisms, fails to form HCOOCH$_3$ both in the gas phase and on the dust, with its abundance being approximately zero, as shown in Figures \ref{fig:gasspec} and \ref{fig:icespec}.

\subsubsection{CH$_2$OHCHO}
CH$_2$OHCHO, an isomer of HCOOCH$_3$, primarily forms on dust grains through radiative mechanisms and is subsequently desorbed into the gas phase via sputtering mechanisms. Therefore, only the MC1 to MC4 models can synthesize CH$_2$OHCHO effectively.
The reaction between CH$_2$OH$^*$ (HCO$^*$) and HCO (CH$_2$OH) on dust grains is the primary formation pathway for JCH$_2$OHCHO.
 When the simulated molecular cloud evolves to 1$\times 10^6$ years,
JCH$_2$OHCHO forms through radiative mechanisms in the MC1 to MC4 models, with 80\% forming in the ice mantle and 20\% on the dust surface. Sputtering desorption is the dominant pathway for the formation of CH$_2$OHCHO, accounting for over 95\%, while photodesorption contributes less than 1\%, and the rest mainly coming from gas-phase dissociative recombination reactions. 
The abundance of CH$_2$OHCHO is directly correlated with the sputtering rate in the models. With all other conditions being equal, a higher sputtering rate results in a higher gas-phase abundance of CH$_2$OHCHO. Consequently, the MC2 model exhibits the highest abundance, followed by MC3, MC1, and MC4. The abundance fluctuations are notably large in the MC1 and MC4 models due to their lower sputtering rates, causing CH$_2$OHCHO that desorbs into the gas phase to be quickly destroyed, leading to significant variability.
In the MC5 model, JCH$_2$OHCHO can be efficiently produced on dust grains. As shown in Figure \ref{fig:icespec}, after approximately $2 \times 10^5$ years, its abundance on the dust grains aligns with the trends observed in the other four models. However, because most of the JCH$_2$OHCHO is trapped within the ice mantle and lacks a sputtering mechanism, these molecules cannot be desorbed into the gas phase. The MC6 model fails to synthesize CH$_2$OHCHO in both the gas phase and on the dust grains.

\subsubsection{CH$_3$CHO}
In all models, CH$_3$CHO primarily forms through gas-phase reactions. In the MC1 to MC4 models, over 75\% of CH$_3$CHO is produced this way, while in the MC5 and MC6 models, gas-phase formation accounts for nearly 100\%. The radiative mechanisms promote the formation of precursor molecules for CH$_3$CHO ( C$_2$H$_5$, ICH$_3^*$/IHCO$^*$ + IHCO/ICH$_3$ $\rightarrow $ IC$_2$H$_5$ ) within the ice mantle, where they freeze and can persist for extended periods. These precursor molecules are then desorbed into the gas phase via sputtering, where they undergo further reactions to form CH$_3$CHO. Although these precursors can also form on the dust surface, they react rapidly with hydrogen atoms and therefore have short lifespans ( gC$_2$H$_5$ + gH $\rightarrow$ gC$_2$H$_6$ with timescales shorter than the desorption timescale).
CH$_3$CHO can only be effectively formed in the MC1 to MC4 models, primarily through gas-phase reactions. In the MC5 and MC6 models, although small amounts of CH$_3$CHO can form, the destruction rate due to ion reactions exceeds the formation rate, leading to the eventual destruction of CH$_3$CHO.

When the simulated molecular cloud evolves to 1$\times 10^6$ years, in the MC1 model, CH$_3$CHO is predominantly formed through gas-phase reactions, with a smaller portion resulting from sputtering desorption of CH$_3$CHO from the ice mantle. The contribution of photodesorption is negligible.
The primary gas-phase reaction pathway is recombination reactions, C$_2$H$_5$OH $\stackrel{\text{H$_3^+$,H$_3$O$^+$,HCO$^+$}}  \longrightarrow$ C$_2$H$_5$OH$_2^+$ $\stackrel{\text{e$^-$}}\rightarrow$ CH$_3$CHO, which account for approximately 27.49\% of the total CH$_3$CHO formation. C$_2$H$_5$OH primarily forms within the dust ice mantle through the newly introduced radiative mechanisms and then desorbs into the gas phase via sputtering.
The second gas-phase reaction pathway is O + C$_2$H$_5$ $\rightarrow$ CH$_3$CHO + H, accounting for approximately 27.10\% of the total formation. 
Notably, C$_2$H$_5$ in the gas phase is desorbed from the dust ice mantle via sputtering.
On the dust, C$_2$H$_5$ primarily forms through two pathways. The first pathway is the reaction on the dust surface, gH + gC$_2$H$_4$ $\rightarrow$ gC$_2$H$_5$
which accounts for 61.58\% of C$_2$H$_5$ formation on the dust.
However, before desorption, gC$_2$H$_5$ rapidly reacts with gH to form gC$_2$H$_6$ on the dust surface, which makes it difficult for gC$_2$H$_5$ to persist.
The second formation pathway occurs via radiative excitation reactions within the dust ice mantle, contributing 38.39\% to the total formation of C$_2$H$_5$ on the dust. The primary radiative excitation reaction involved is ICH$_2^*$/ICH$_3^*$ + ICH$_3$/ICH$_2$ $\rightarrow$ IC$_2$H$_5$. Most of the C$_2$H$_5$  formed in this process become trapped within the ice mantle and remain stable until sputtering desorbs them into the gas phase. Models incorporating radiative mechanisms show C$_2$H$_5$ abundances on dust that are approximately three orders of magnitude higher than those in the MC6 model, which lacks radiative mechanisms. 
In the MC1 to MC4 models, the gas-phase abundance of C$_2$H$_5$ ranges from approximately $1 \times 10^{-11}$ to $4 \times10^{-10}$ while in the MC5 and MC6 models, which lack the sputtering mechanism, C$_2$H$_5$ does not form in the gas phase.

In the MC2 model, CH$_3$CHO begins forming in the gas phase around $5 \times 10^4$ years, with its abundance gradually increasing over time. 
When the simulated molecular cloud evolves to 1$\times 10^6$ years,
its abundance reaches a peak of approximately $1 \times 10^{-9}$, the highest among all models. This is because the sputtering rate in the MC2 model is the highest, with the ice mantle sputtering rate being 4 to 5 times greater than that of the MC1 model. This higher rate allows more species to desorb from the dust ice mantle into the gas phase. Consequently, at $1 \times 10^6$ years, the gas-phase abundance of C$_2$H$_5$ in the MC2 model is approximately 8 times higher than in the MC1 model, and the abundance of C$_2$H$_5$OH is about 6 times higher, leading to an increased abundance of CH$_3$CHO in the gas phase.
In the MC2 model, the recombination reaction of C$_2$H$_5$OH$_2^+$ is the primary formation pathway for CH$_3$CHO, accounting for approximately 30.51\%. The second major pathway, the reaction between gas-phase O and C$_2$H$_5$, contributes around 25.35\%, while the sputtering desorption of CH$_3$CHO from the ice mantle into the gas phase accounts for about 22.48\%. The primary formation channels of gas-phase C$_2$H$_5$ and C$_2$H$_5$OH$_2^+$ in the MC2 model are consistent with those in the MC1 model.

The evolutionary trend of CH$_3$CHO in the MC3 model is almost identical to that in the MC2 model, as depicted by the solid green line in Figure \ref{fig:gasspec}. However, the abundance of CH$_3$CHO in the MC3 model is slightly lower, with a difference within a factor of 2. CH$_3$CHO primarily forms through several pathways: the reaction between gas-phase O and C$_2$H$_5$, which accounts for approximately 26.18\% of its formation, the sputtering desorption of CH$_3$CHO from the ice mantle into the gas phase, contributing about 22.41\%, and the recombination reaction involving C$_2$H$_5$OH$_2^+$, which contributes around 29.23\%. The formation channels of gas-phase C$_2$H$_5$ and C$_2$H$_5$OH$_2^+$ in the MC3 model are consistent with those in the MC2 model. 
Gas-phase C$_2$H$_5$ primarily originates from sputtering desorption of C$_2$H$_5$ frozen within the ice mantle, while C$_2$H$_5$OH$_2^+$ is mainly produced from the ion-induced destruction of C$_2$H$_5$OH.

Among the four models that include sputtering and radiative mechanisms, the MC4 model has the lowest CH$_3$CHO abundance, reaching its peak at around $7 \times 10^{-11}$ at $9 \times 10^5$ years, as shown by the red solid line in Figure \ref{fig:gasspec}, which is approximately one order of magnitude lower than the MC2 model. The sputtering rate in the MC4 model is the lowest, and after $5 \times 10^4$ years, the sputtering rate within the ice mantle remains nearly constant, as indicated by the yellow dashed line in Figure \ref{fig:sputtering}. In contrast, the sputtering rates in the MC2 and MC3 models continue to gradually increase over time.
As a result, the abundance of CH$_3$CHO, C$_2$H$_5$, and C$_2$H$_5$OH desorbed from the ice mantle into the gas phase via sputtering is lower in the MC4 model than in the other models. The evolution trend and primary formation pathways of CH$_3$CHO in the MC4 model are similar to those in the MC1 model.

In the MC5 and MC6 models, a small amount of CH$_3$CHO forms in the gas phase after $3 \times 10^5$ years, reaching a maximum abundance of approximately $8 \times 10^{-12}$. However, after $8 \times 10^5$ years, all CH$_3$CHO is destroyed by ions. The primary formation and destruction pathways are as follows: 
H$_3$O$^+$ + C$_2$H$_2$ $\rightarrow$ C$_2$H$_5$O$^+$ $\stackrel{\text{e$^-$}}\longrightarrow$  CH$_3$CHO $\stackrel{\text{ions}}\longrightarrow$ C$_2$H$_5$O$^+$.

However, in the MC5 model, molecules like CH$_3$CHO, C$_2$H$_5$, and C$_2$H$_5$OH can form within the ice mantle, with abundances consistent with those in the other four models, as shown in Figure \ref{fig:icespec}. Due to the lack of sputtering mechanisms, these molecules within the ice cannot desorb into the gas phase, preventing them from contributing to the formation of CH$_3$CHO in the gas phase. In the MC6 model, which lacks radiative mechanisms, CH$_3$CHO and its related precursor molecules, such as C$_2$H$_5$ and C$_2$H$_5$OH, cannot form within the ice mantle either.

\subsubsection{CH$_3$OCH$_3$}
CH$_3$OCH$_3$ predominantly forms within the dust ice mantle through radiative excitation mechanisms and is subsequently desorbed into the gas phase via sputtering. As a result, CH$_3$OCH$_3$ can only form in the gas phase in the MC1 to MC4 models. In the MC5 model, CH$_3$OCH$_3$ is formed within the dust ice mantle but cannot be desorbed into the gas phase due to the absence of sputtering mechanisms. The MC6 model fails to produce CH$_3$OCH$_3$ both in the dust mantle and in the gas phase.

In the MC1 to MC4 models, the primary formation pathway and contribution percentages are nearly identical. The most significant pathway, where CH$_3$OCH$_3$ is desorbed from the dust ice mantle into the gas phase via sputtering, accounts for approximately 74\% of CH$_3$OCH$_3$ formation. Within the ice mantle, the primary radiative excitation reactions responsible for CH$_3$OCH$_3$ formation are ICH$_3^*$(ICH$_3$) + ICH$_3$O(ICH$_3$O$^*$) $\rightarrow$ ICH$_3$OCH$_3$, contributing about 81\%, and ICH$_2^*$(ICH$_2$) + ICH$_3$OH(ICH$_3$OH$^*$) $\rightarrow$  ICH$_3$OCH$_3$, contributing approximately 17\%. In the gas phase, the formation of CH$_3$OCH$_3$ primarily occurs through a recombination process involving the dissociation and recombination of CH$_3$OCH$_3$ desorbed from the dust (JCH$_3$OCH$_3$ $\rightarrow$ CH$_3$OCH$_3$ $\stackrel{\text{ions}}\longrightarrow$ CH$_3$OCH$_4^+$ $\stackrel{\text{$e^{-}$}}\longrightarrow$ CH$_3$OCH$_3$), which accounts for around 26\%.

The main difference among the MC1 to MC4 models lies in the timing and abundance of CH$_3$OCH$_3$ formation, which is directly proportional to the sputtering rate. In the MC2 and MC3 models, CH$_3$OCH$_3$ synthesis begins effectively around $4 \times 10^4$ years, with its abundance gradually increasing over time. The abundance and evolution trends of CH$_3$OCH$_3$ in these two models are quite similar. In contrast, CH$_3$OCH$_3$ formation in the MC1 and MC4 models occurs later, at around $6 \times 10^4$ and $1 \times 10^5$ years, respectively, with lower abundances than in the MC2 and MC3 models. The MC2 model exhibits the highest peak abundance at approximately $5 \times 10^{-9}$, more than an order of magnitude higher than the MC4 model.

As shown in Figure \ref{fig:icespec}, CH$_3$OCH$_3$ forms effectively on the dust in all models from MC1 to MC5, with similar abundances. However, in the MC5 model, due to the lack of sputtering mechanisms, CH$_3$OCH$_3$ cannot desorb from the ice mantle into the gas phase, resulting in a near-zero gas-phase abundance. The MC6 model fails to form CH$_3$OCH$_3$ in both the dust mantle and the gas phase.

\subsubsection{C$_2$H$_5$OH}
C$_2$H$_5$OH is an isomer of CH$_3$OCH$_3$, and it shares a similar formation mechanism. It mainly forms in dust ice mantles through radiative excitation reaction, and then desorbs into the gas phase via sputtering. 
C$_2$H$_5$OH forms efficiently only in the MC1 to MC4 models. In the MC2 model, its abundance peaks at approximately $3 \times 10^{-9}$ at 1$\times 10^6$ years, while in the MC4 model, the abundance is lower, around $2 \times 10^{-10}$. In these models, 99\% of C$_2$H$_5$OH in the gas phase produced by sputtering desorption from the dust ice mantles. Within the ice mantles, C$_2$H$_5$OH mainly forms through radiative mechanisms, primarily via the reaction ICH$_3^*$(ICH$_3$) + ICH$_2$OH(ICH$_2$OH$^*$) $\rightarrow$ IC$_2$H$_5$OH, which accounts for about 82\% of its formation. The secondary pathway, ICH$_2^*$(ICH$_2$) + ICH$_4$O(ICH$_4$O$^*$) $\rightarrow$ IC$_2$H$_5$OH, contributes approximately 17\%. Photodesorption is negligible across all four models, contributing less than 0.1\%. In the MC5 and MC6 models, a small amount of C$_2$H$_5$OH can form around $4 \times 10^5$ years, mainly through limited ion-molecule recombination reactions in the gas phase, but it is rapidly dissociated. The MC5 model can produce C$_2$H$_5$OH in the ice mantles.

Our model can trap species within the ice mantle, where photons penetrate the partially active ice layers and photodissociate bulk ice species, producing radicals.
Species in the ice mantle can become excited through radiative excitation and react with nearby species, leading to the formation of  COMs. The sputtering mechanism desorbs these COMs, along with other molecules like C$_2$H$_5$, from the ice mantle into the gas phase. C$_2$H$_5$ is a crucial precursor for the formation of CH$_3$CHO in the gas phase. The sputtering rate directly influences the abundance of COMs in the gas phase; a higher sputtering rate increases the desorption of molecules from the ice mantle, resulting in a greater gas-phase abundance. Furthermore, the sputtering mechanism modifies the species abundance distribution between the dust and gas phases, enhancing chemical reactions in the gas phase.
These COMs exhibit competitive relationships. For example, HCO is a precursor molecule for the formation of CH$_2$OHCHO, HCOOCH$_3$, and CH$_3$CHO, leading to competition in their formation pathways. Additionally, there are synergistic relationships among these species. For instance, in the gas phase, C$_2$H$_5$OH$_2^+$ is primarily produced by the ion-induced destruction of C$_2$H$_5$OH. The recombination reaction of C$_2$H$_5$OH$_2^+$ with $e^-$ can form CH$_3$CHO, which accounts for about 30\% of the CH$_3$CHO formation in the MC1 to MC4 models. In all models, the cosmic-ray-induced grain heating desorption mechanism is ineffective for the desorption of COMs that we are interested in.

\subsubsection{COMs on Grain}
\begin{figure*}
	\includegraphics[width=\columnwidth]{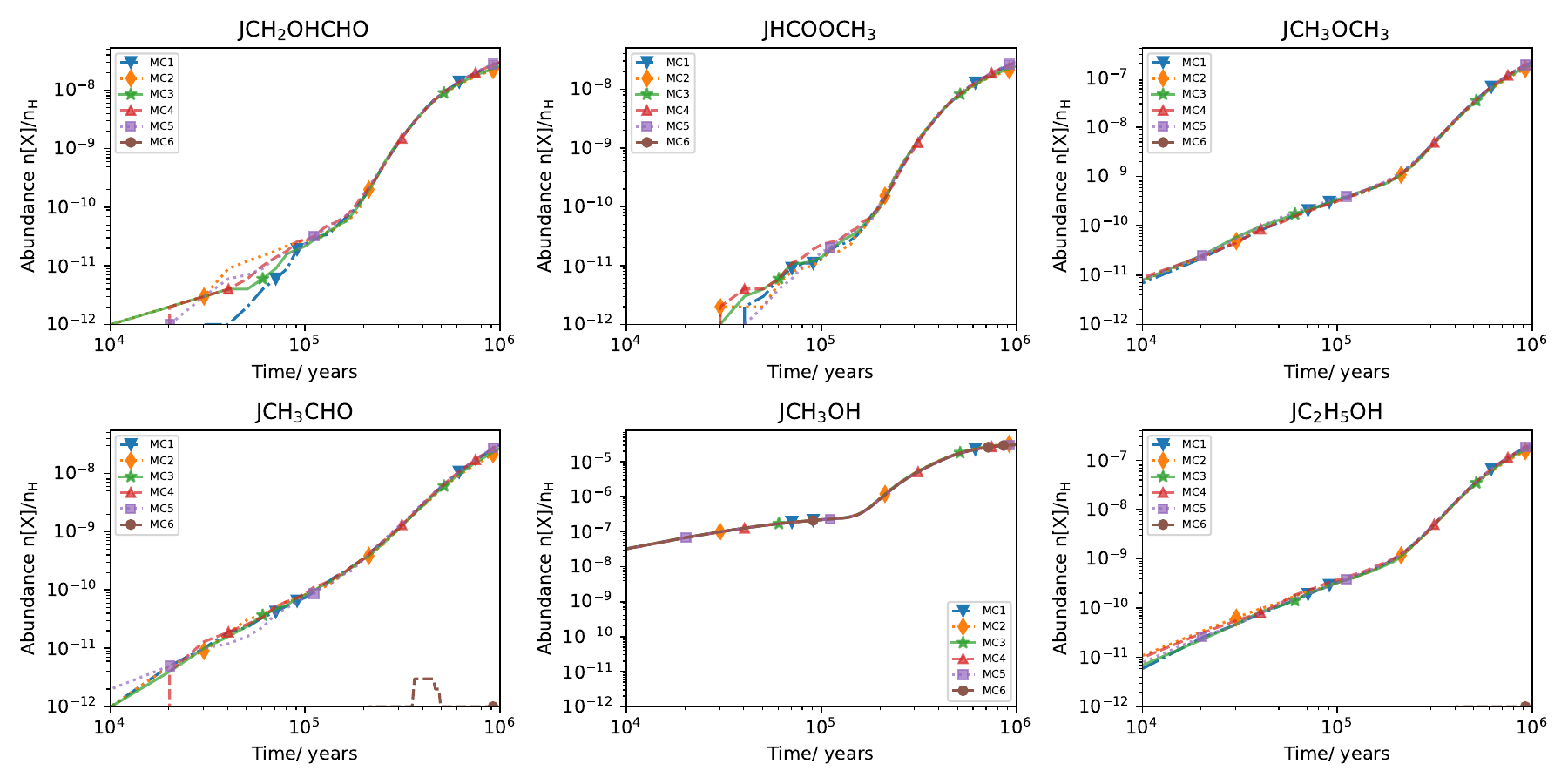}
    \caption{Abundance Ratios of Dust Species Relative to H in Different Models.}
    \label{fig:icespec}
\end{figure*}

Figure \ref{fig:icespec} displays the relationship between the abundance of O-bearing COMs within dust and time across various models. 
It demonstrates that models MC1 to MC5 efficiently synthesized these O-bearing COMs, and their abundances remaining relatively consistent.
This consistency can be attributed to the presence of the radiolysis mechanism in the MC1 to MC5 models.
The formation of COMs was attributed to cosmic rays interacting with dust, leading to radiolysis reactions that excited the species. These excited species could then react with neighboring species, without the need for diffusion, to produce COMs. The channels mainly formed by these COMs are detailed in Table \ref{tab:mainchannels}.
The initiation of COMs formation on dust occurred around $10^4$ years when several layers of ice had formed.
The abundance of COMs and the number of ice layers increased over time.
However, due to the absence of sputtering desorption mechanisms, COMs in the dust ice mantle of model 5 could not desorb into the gas phase. 
In the case of the MC6 model, it could not synthesize these COMs, except for CH$_3$OH, even within the ice layers, as it lacked radiolysis mechanisms.

\begin{figure*}
	\includegraphics[width=\columnwidth]{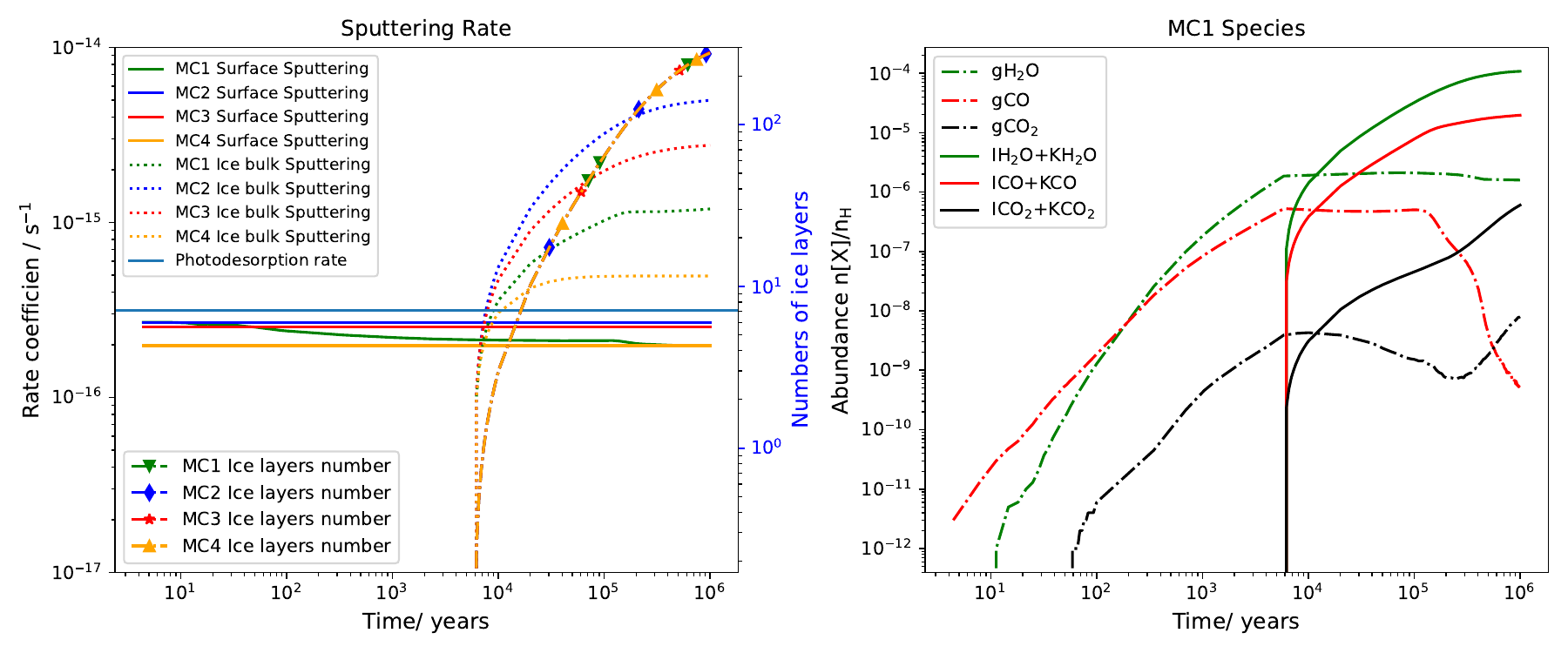}
    \caption{Left panel: Dust surface and ice mantle sputtering rate coefficients versus time for different models. Right panel (MC1 model): The abundances of H$_2$O, CO, and CO$_2$ molecules change with time.}
    \label{fig:sputtering}
\end{figure*}

The left panel of Figure \ref{fig:sputtering}  shows the sputtering rate coefficients of the dust surface and bulk ice, along with the variations in the number of ice layers over time for different models.
In the right panel of Figure \ref{fig:sputtering}, it shows the fractional abundances of grain species with respect to hydrogen in the MC1 model.
It is evident that the dust surface sputtering rate coefficients of MC2-MC4 remain constant over time, as they are determined by the sputtering rate coefficients of a grain based on a single component in these three models. Additionally, the bulk ice sputtering rate coefficients are solely dependent on the number of ice layers and can be computed using equation \ref{eq:eff}. 
The left panel of Figure \ref{fig:sputtering} shows that the sputtering rate coefficient of CO ices is the highest, followed by CO$_2$ ices, and finally H$_2$O ices.
Both the surface and bulk ice sputtering rate coefficients of MC1 were influenced by the abundance of CO, CO$_2$, and H$_2$O on the grain.
The left graph in Figure \ref{fig:sputtering} shows that the sputtering rate coefficient of the surface in MC1 was initially equal to that of MC2 when the time was less than 10 years. 
However, over time, it gradually decreased and eventually became comparable to the sputtering rate coefficient of MC4 after approximately $3 \times 10^5$ years.
The reason for this reduction is evident from the dash-dot line in the right graph of Figure \ref{fig:sputtering}. Before $10^2$ years in the model, the surface of MC1 had lower abundances of CO$_2$ and H$_2$O, and its sputtering rate was primarily influenced by CO. As time passed, the abundance of H$_2$O gradually increased, leading to a gradual decrease in the sputtering rate coefficient. After 10$^5$ years, the abundance of CO dropped sharply(red dashdot line), while the abundance of H$_2$O remained significantly higher (two to three orders of magnitude) compared to that of CO and CO$_2$. Moreover, the abundance of H$_2$O reached a stable state. As a result, the sputtering rate at this stage was primarily influenced by H$_2$O.
The sputtering rate coefficient of MC1 bulk ice was influenced by the number of ice layers and the abundance of H$_2$O, CO, and CO$_2$ on ice mantle. Ice layer formation in MC1 began approximately 1$\times 10^4$ years. Over a short period of time, the number of ice layers increased sharply, reaching 60 layers in approximately 1$\times 10^5$ years.
The sputtering rate coefficient gradually increased from 2$\times 10^4$ years and stabilized approximately after 1$\times 10^5$ years. This is because the influence of the number of ice layers diminishes gradually after 1$\times 10^5$ years. At the same time, the abundance of CO within the ice stabilized (solid red line on the right in Figure \ref{fig:sputtering}), 
while  the abundance of H$_2$O and CO$_2$ increased slowly, and their impact on the sputtering rate was smaller than that of CO.
As a result, the sputtering rate of the bulk ice remains nearly constant after 1$\times 10^5$ years.

\begin{figure*}
	\includegraphics[width=\columnwidth]{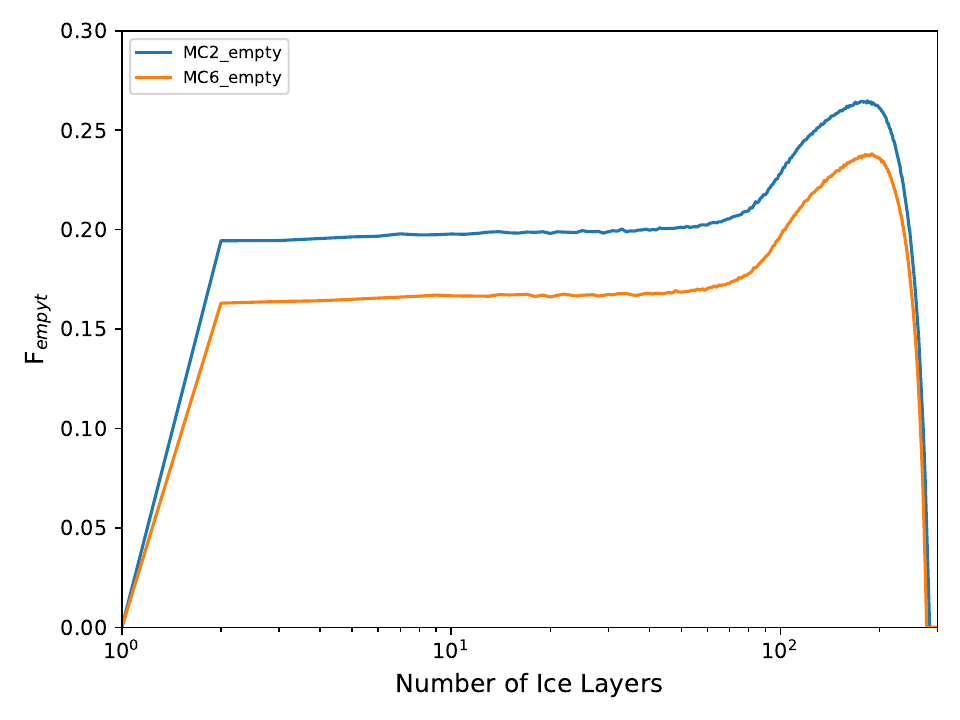}
    \caption{The distribution of F$_{empty}$ in different ice layers of the MC2 and MC6 models at 10$^6$ years.}
    \label{fig:empty}
\end{figure*}

Figure \ref{fig:empty} shows the distribution of the fraction of empty normal sites (F$_{empty}$) across different ice layers. Here, the ``1'' on the x-axis represents the deepest ice layer, while increasing values correspond to layers closer to the ice surface. The fraction of empty normal sites is defined as F$_{empty}$ = Total$_{empty}$/N$_s$, where Total$_{empty}$ is the population of empty normal sites in the bulk of ice. These empty are mainly caused by photodissociation or sputtering normal sites species. 

The distribution of  F$_{empty}$ in the MC2 and MC6 models is nearly identical in the bottom 100 ice layers, with proportions of approximately 20\% and 16\%, respectively. Beyond the 100th layer, F$_{empty}$ gradually increases, 
reaching a maximum at around the 200th layer, with values of approximately 26\% and 23\% for the MC2 and MC6 models, respectively. Above the 250th layer, F$_{empty}$ begins to decrease, with proportions falling below 15\%. This decline is due to the relatively late formation of the topmost ice layers, resulting in shorter exposure to photodissociation and sputtering. In the MC2 model, F$_{empty}$ is consistently 2–4\% higher than in the MC6 model within the first 250 layers, primarily due to the sputtering mechanism. Beyond this depth, the differences between the two models become less pronounced as F$_{empty}$ begins to decrease in the uppermost layers. Sputtering has its most significant impact on the ice layers between 90 and 240, where a higher number of species desorb into the gas phase compared to other layers.

\subsubsection{reactive desorption}
\begin{figure}
	\includegraphics[width=\columnwidth]{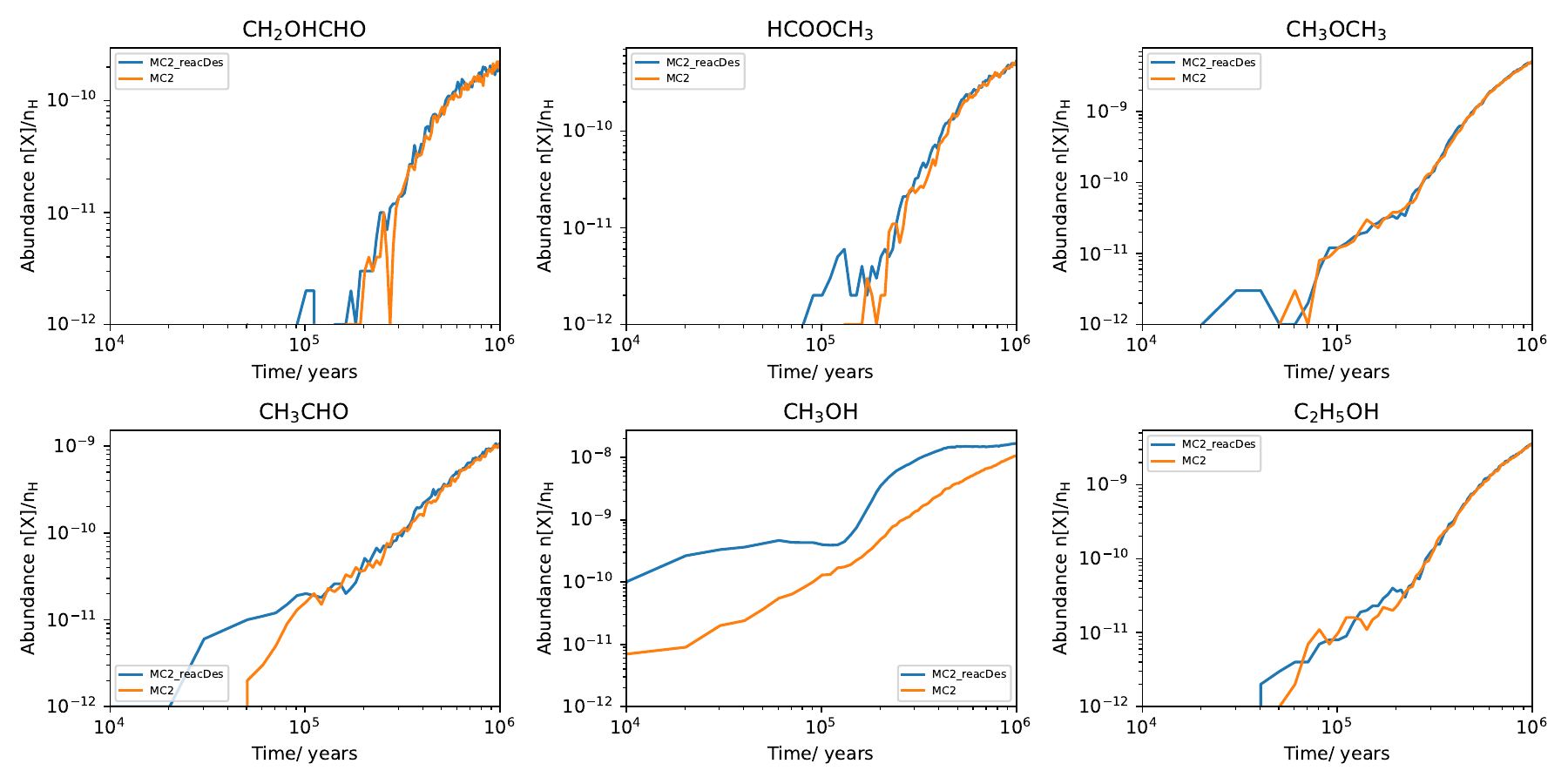}
    \caption{Abundance Ratios of Gas Species Relative to H in MC2 and MC2\_reacDes Models.}
    \label{fig:gas_reacdes_spec}
\end{figure}

A number of studies have reported the importance of reactive desorption for the formation of COMs \citep{2007AA...467.1103G,2017MolAs...6...22W,2020ApJS..249...26J}. Therefore, we incorporated the reactive desorption mechanism into the MC2 model, referred to as the MC2\_reacDes model. 
According to the work of \cite{2016MNRAS.459.3756R}, we set the reactive desorption efficiency for all chemical reactions on the dust surface to 1\% in the MC2\_reacDes model. 
Figure \ref{fig:gas_reacdes_spec} compares the results of the MC2 model and the MC2\_reacDes model. It can be seen that reactive desorption significantly affects the abundance of CH$_3$OH, while its impact on other COMs is relatively minor. This is because CH$_3$OH primarily forms on the dust surface and can desorb into the gas phase via the reactive desorption mechanism. 
In contrast, other COMs mainly form within the ice mantles, or their precursor molecules desorb from the ice mantle into the gas phase for further formation, where reactive desorption does not occur within the ice mantles.

In the MC2 model, the primary pathway for gas-phase CH$_3$OH formation is photodesorption CH$_3$OH from the dust surface. However, in the MC2\_reacDes model, reactive desorption becomes the dominant pathway for CH$_3$OH desorption into the gas phase. Since the efficiency of reactive desorption is higher than that of photodesorption, the abundance of CH$_3$OH in the MC2\_reacDes model is higher than that in the MC2 model. For other COMs, the inclusion of the reactive desorption mechanism allows a small amount of these molecules to desorb into the gas phase at earlier times. As time progresses, sputtering of COMs from the ice mantles becomes the dominant process. As a result, the abundance evolution of these COMs shows only minor differences between the MC2 and MC2\_reacDes models at earlier times, while their abundances nearly overlap at later times.

\subsection{Other}
In addition to observed COMs, TMC-1 has also been observed to contain simple species and carbon chain species. 
To investigate whether the introduced radiation and sputtering mechanisms affected the formation of other gas phase reactions, such as carbon-chain molecules, we analyzed the chemical evolution processes of the hydrocarbon family, cyanopolyyne family, and methylcyanopolyynes family.

\begin{figure*}
	\plotone{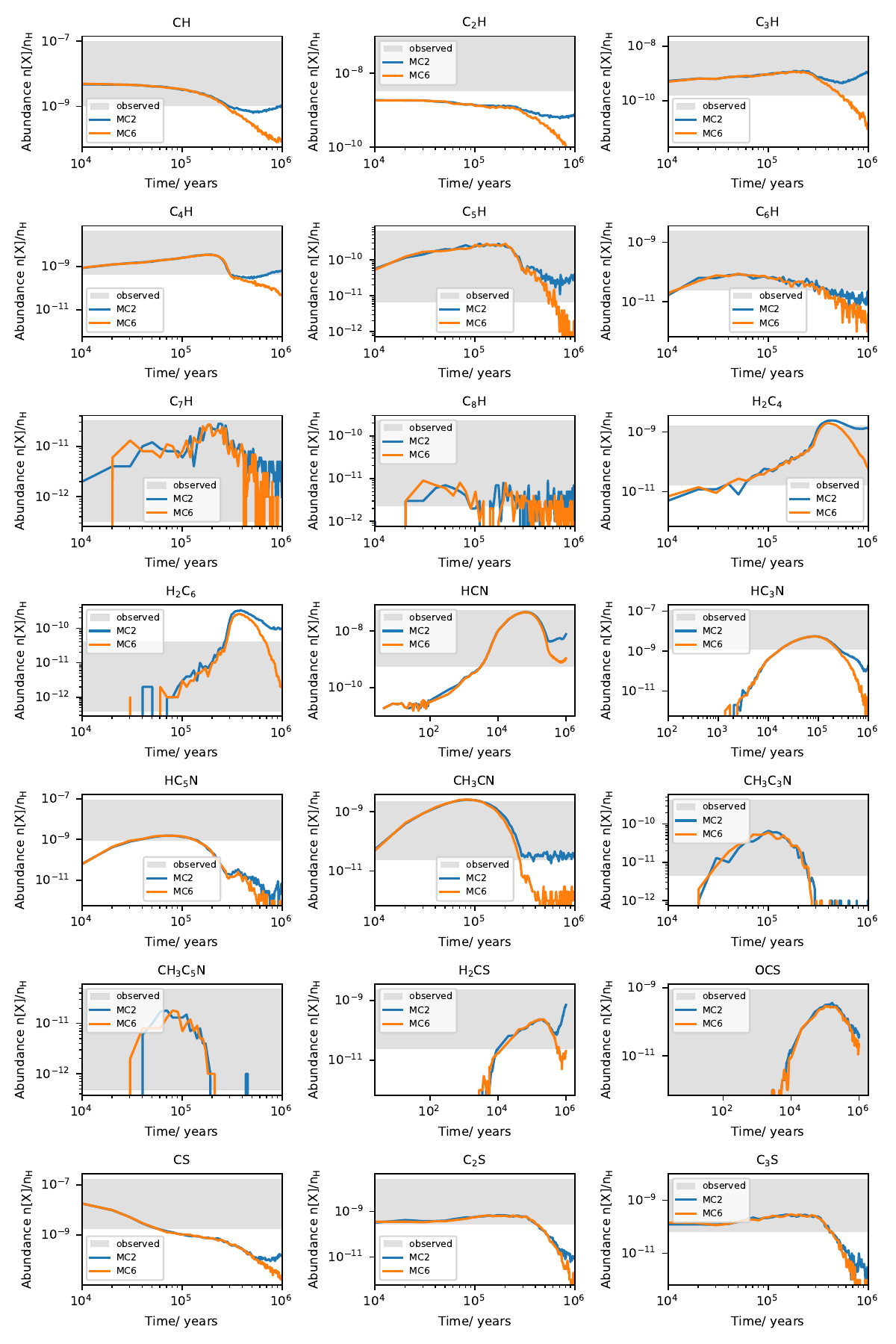}
    \caption{Abundance Ratios of Gas Carbon Chain Species Relative to H in Model 1 and Model 6.}
    \label{fig:carbonone}
\end{figure*}

Figure \ref{fig:carbonone} shows the abundance of gas phase carbon chain species in the MC2 and MC6 models.
The gray area represents a range of one order of magnitude above and below the observed values.
Most hydrocarbons, nitrogen-containing carbon chain molecules, and sulfur-containing carbon chain molecules are in good agreement with observed values, typically within one order of magnitude.
These molecules could be synthesized in both MC2 and MC6 since their primary formation environment was in the gas phase. Therefore, even in MC6, which lacked cosmic ray non-thermal mechanism, these molecules could still be efficiently synthesized. 

In our models, hydrocarbons (C$_\text{n}$H, where n = 1, 2, 3, ...) were synthesized early (around 10$^4$ years), primarily through reactions described in equations \ref{eq:carb1} and \ref{eq:carb2}, as well as other recombination reactions such as C$_n$H$_3^+$ + e$^-$ $\rightarrow$ C$_n$H + H$_2$ and C$_{n+1}$H$^+$ + e$^-$ $\rightarrow$ C$_n$H + C, before ionic (C$^+$) or atomic (C) carbon became locked into CO or absorbed onto dust.
On average, more than 20 pathways contributing to the generation of each hydrocarbon, with the mentioned pathways collectively accounting for approximately 60\% of the total production channels.
Figure \ref{fig:carbonone} shows that the MC2 model, which included radiation and sputtering mechanisms, yielded synthesis results for the hydrocarbon family nearly identical to those of the MC6 model, which lacked both mechanisms. 
In the MC2 and MC6 models, the abundances of C$_\text{n}$H nearly overlap before 3 $\times$ 10$^5$ years. After 3 $\times$ 10$^5$ years, the abundance of C$_\text{n}$H in the MC6 model decreases sharply, while in the MC2 model, the abundance of C$_\text{n}$H stabilizes or even slowly increases, as seen with C$_3$H. 
Before $10^4$ years, C$_\text{n}$H is produced rapidly in large quantities. As ionic (C$^+$) or atomic (C) carbon becomes locked into CO, the efficiency of C$_\text{n}$H formation decreases. Concurrently, the destruction rate of C$_\text{n}$H exceeds its production rate, resulting in a sharp decline in its abundance after 3$\times$10$^5$ years. 
However, in the MC2 model, which incorporates sputtering and radiative mechanisms, C$_\text{n}$H and other species that serve as precursor molecules for C$_\text{n}$H can desorb from dust into the gas phase after 3$\times$10$^5$ years.
Between 3$\times$10$^5$ and 1$\times$10$^6$ years, this desorption via the sputtering mechanism becomes an important formation pathway for C$_\text{n}$H. 
For example, during this period, in the MC2 model, 36.9\% of the C$_3$H in the gas phase is formed through the sputtering mechanism that desorbs C$_3$H from the ice mantle. Additionally, about 43.05\% of the C$_3$H in the gas phase is produced from reactions involving C$_3$ that is sputtered from the ice mantle (JC$_3$ $\rightarrow$ C$_3$ $\stackrel{\text{H$_3^+$,H$_3$O$^+$,HCO$^+$}}  \longrightarrow$ C$_3$H$^+$ $\stackrel{\text{H$_2$}}  \longrightarrow$ C$_3$H$_3^+$ $\stackrel{\text{e$^-$}}  \longrightarrow$ C$_3$H). At 1$\times$10$^6$ years, the abundance of C$_3$ in the gas phase is two orders of magnitude higher in the MC2 model compared to the MC6 model.
As a result, the introduction of these new mechanisms after 3$\times$10$^5$ years increases the gas-phase abundance of C$_\text{n}$H. At the final time (1$\times$10$^6$ years), the abundance of C$_\text{n}$H in the MC2 model is higher than in the MC6 model by one to two orders of magnitude.

We investigated the efficiency of the three primary formation mechanisms (\ref{eq:cyanopolyynef1_1}, \ref{eq:cyanopolyynef2_1}, \ref{eq:cyanopolyynef3_1}) of the cyanopolyyne family in our models.
Research by \citet{2018MNRAS.474.5068B} showed that the primary production mechanism for larger cyanopolyynes molecules (HC$_n$N, where n $\geq$ 5) was Mechanism 1, while Mechanism 2 could be disregarded, and Mechanism 3 was inefficient.
Mechanism 2 was effective in synthesizing smaller cyanopolyyne molecules. 
Table \ref{tab:CyanopolyynePro} shows the formation probability of cyanopolyyne family channels in MC2 model at 1$\times 10^6$ years. In our model, Mechanism 1 was the primary formation channel for cyanopolyyne family species, whether they were large or small cyanopolyyne molecules. 
In Mechanism 1, H$_2$C$_n$N$^+$ is mainly formed through a series of ion-molecule reactions, as shown in Formula \ref{eq:Mech_1}.
\begin{eqnarray}
\rm C/C^+  \rightarrow \cdots \rightarrow C_n^+ \rightarrow \cdots  \rightarrow C_nN^+ \stackrel{H_2} \longrightarrow HC_nN^+ \stackrel{H_2} \longrightarrow H_2C_nN^+ \stackrel{e^-} \longrightarrow HC_nN \label{eq:Mech_1}
\end{eqnarray}
The efficiency of Mechanism 2 and Mechanism 3 was negligible.
These cyanopolyyne molecules had peak modeled abundances around 1$\times 10^5$ years. 
Figure \ref{fig:carbonone} shows that both the radiolysis mechanism and the sputtering mechanism have no significant impact on the synthesis of the cyanopolyyne family in the gas phase before 3$\times$10$^5$ years.
Before 3$\times$10$^5$ years, the abundance evolution curves of HC$_n$N in the MC2 and MC6 models almost completely overlap. After 3$\times$10$^5$ years, the abundance of HC$_n$N in the MC6 model declines rapidly, whereas in the MC2 model, it decreases more slowly or tends to stabilize. At 1$\times$10$^6$ years, the maximum abundance difference of HC$_n$N between the two models is about two orders of magnitude. The MC2 model, which includes new mechanisms, allows more species to desorb from the dust ice mantle into the gas phase, enhancing the formation of HC$_n$N in the gas phase. For example, the abundances of CN, C$_2$H$_2$, and C$_3$H$_3$ in the gas phase are several to hundreds of times higher in the MC2 model than in the MC6 model. This increase facilitates the formation of HC$_3$N in the gas phase through the following pathways:

\begin{eqnarray}
\rm CN + C_2H_2  \rightarrow HC_3N + H, \label{eq:hc3n_1}\\
\rm N + C_3H_3  \rightarrow  HC_3N + H_2. \label{eq:hc3n_2}
\end{eqnarray}

\begin{deluxetable*}{lllllc}
\tablecaption{Cyanopolyyne Family Channel Formation Probability in MC2 at 1$\times 10^6$ years.}
\tabletypesize{\scriptsize}
\label{tab:CyanopolyynePro}
\tablehead{\colhead{Species} &\colhead{Mechanism 1} & \colhead{Mechanism 2}& \colhead{Mechanism 3}&
\colhead{Other} &\colhead{Total Pathways}
}
\startdata
      HCN & 60.24\% & -&-&39.76\%&73\\
      HC$_3$N & 48.93\% & 4.84\%&0.0\%&46.23\%&25\\
      HC$_5$N & 96.29\% & 0.95\%&0.50\%&2.26\%&14\\
      HC$_7$N & 76.68\% & 1.81\%&1.19\%&20.32\%&13\\
      HC$_9$N & 70.63\% & 1.94\%&0.65\%&26.78\%&10\\
\enddata
\end{deluxetable*}

The abundance and evolutionary trend of the methylcyanopolyynes family are influenced by HC$_n$N, as HC$_n$N acts as a precursor molecule in the primary formation pathway of CH$_3$C$_n$N. The formation pathway is HC$_n$N $\rightarrow$ CH$_3$C$_n$N$^+$ $\rightarrow$ CH$_3$C$_{n}$N.

Simple sulfur-bearing carbon chains, such as C$_2$S and C$_3$S, were also efficiently synthesized, consistent with observations occurring in approximately 10$^5$ years.
Our simulation results indicate that both the radiolysis and sputtering mechanisms have no significant impact on the formation of gas-phase carbon chain molecules before 3$\times$10$^5$ years.
After 3$\times$10$^5$ years, the MC2 model, which includes new mechanisms, is able to significantly increase the abundance of these carbon chain molecules in the gas phase. However, for some larger carbon chain molecules, such as HC$_5$N and CH$_3$C$_5$N, the increase in abundance is relatively small.
Nevertheless, our model still efficiently synthesized these carbon chain molecules in the early stages, aligning with observations.

The evolution trend of these carbon chain molecules was generally consistent, with their abundance gradually decreasing after reaching the peak of synthesis in the early stages. However, during the later stages, this decline was primarily due to their reactions with ions in the gas phase, such as H$_3$O$^+$, H$^+$, and H$_3^+$, as well as reactions with atoms (O, N, etc.). 
Additionally, some of these molecules were also adsorbed onto dust grains. In the later stage, 
ionic (C$^+$) or atomic (C) carbon was locked into CO, and the formation efficiency of these carbon chain molecules was lower than the destruction efficiency. This imbalance was also the reason for the decrease in abundance in the later stage.
However, the model with the newly introduced mechanisms can enhance the formation of carbon chain molecules in the later stages by desorbing molecules from the dust ice mantles. Figure \ref{fig:carbonone_reades} in Appendix \ref{appendix A} shows the impact of reactive desorption on gas-phase carbon-chain molecules. By comparing the MC2 model with the MC2\_reacDes model, we found that reactive desorption has a minimal effect on the abundance of gas-phase carbon-chain molecules. Their abundance evolution trends remain consistent and nearly overlap.

\begin{figure*}
	\plotone{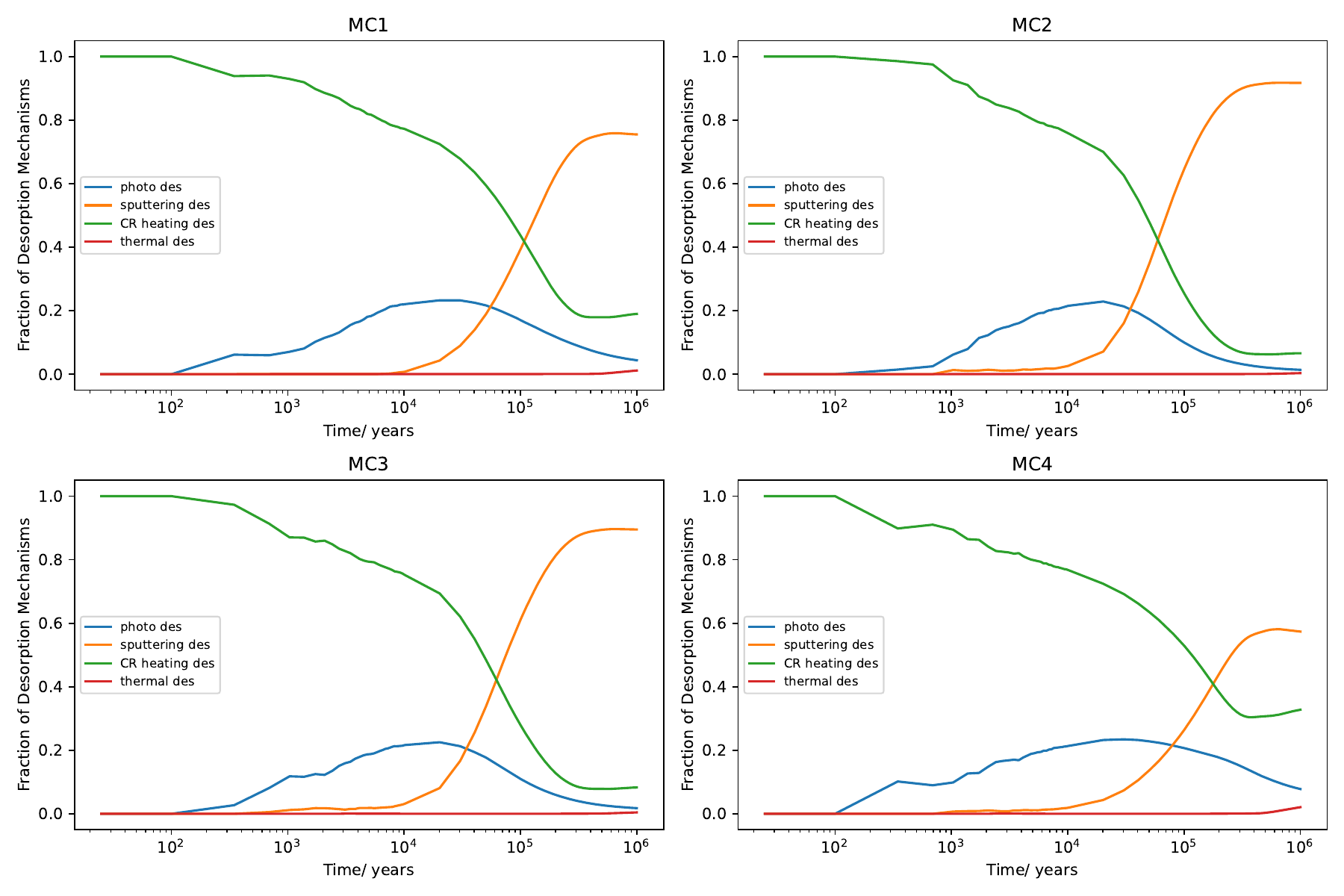}
    \caption{Cumulative Probability of Different Desorption Mechanisms in Various Models.}
    \label{fig:FractionofDesorption}
\end{figure*}

Figure \ref{fig:FractionofDesorption}  shows the cumulative probability of the three non-thermal desorption mechanisms, as well as thermal desorption, in our models over time for all species on grains under cold cloud core conditions.
The red line represents thermal desorption probability, which is highly inefficient under cold source conditions and can be disregarded, accounting for only 0.16\%, 0.01\%, 0.03\%, and 0.94\% in the MC1, MC2, MC3, and MC4 models, respectively, at their respective local best-fitting time (The local best-fitting time is defined in Section \ref{COMs OBS} and listed for different models in Table \ref{tab:obesfit}). 
The blue line represents the photo desorption probability, which accounts for approximately 6.55\%, 3.39\%, 3.27\%, and 9.70\% in the MC1, MC2, MC3, and MC4 models, respectively.
The orange line represents the sputtering desorption probability, with its primary impact occurring approximately $\sim10^4$ years later when the ice layer begins to form, leading to a significant increase in the sputtering rate. Its efficiency accounts for approximately 75.39\%, 89.43\%, 88.80\%, and 58.11\% in the MC1, MC2, MC3, and MC4 models, respectively.
The green line represents the cosmic rays heating desorption probability, initially dominant. However, as sputtering desorption commences, its efficiency gradually diminishes, accounting for approximately 17.9\%, 7.17\%, 7.90\%, and 31.22\% in the MC1, MC2, MC3, and MC4 models, respectively.
The sputtering desorption mechanism can desorb species on both the dust surface and within the ice mantle, whereas the other three desorption mechanisms can only desorb molecules on the dust surface.
When the molecular cloud reaches an age of $\sim 1\times 10^5$ years, the dominant desorption mechanisms are cosmic rays sputtering.

\subsection{Comparison with Observations and Previous Models}
\label{COMs OBS}
We defined the local best-fitting time as the moment when  D(t) reaches its minimum, using the following formula \citep{2006AA...451..551W,2017MNRAS.469..435V},

\begin{eqnarray}
	D(t) = \frac{1}{N_{obs}}\sum_{i}\mid\log[n(X)_{obs,i}]-\log[n(X){_{mod,i}}(t)].
	\label{eq_best_local}
\end{eqnarray}

here, $n(X){_{obs,i}}$ and $n(X){_{mod,i}}(t)$ represent the observed abundances of the $i$-th species and the modeled abundances at time $t$, respectively. $N_{obs}$ denotes the total count of observed species.
For the local best-fitting time, we only consider the five observed COMs: HCOOCH$_3$, CH$_3$OCH$_3$, CH$_3$CHO, CH$_3$OH, and C$_2$H$_5$OH.

Table \ref{tab:obesfit} shows the observed abundance of selected oxygen-bearing COMs in TMC-1, along with the simulated abundance at the local best-fitting time in different models. The last seven rows of the table summarize the results from previous simulations.
The methods employed in various studies are listed in the last column of the table.
While the simulation results from previous models were several orders of magnitude lower than the observed results (indicated by the bold black data in the table \ref{tab:obesfit}), indicating that these COMs could not be effectively synthesized. 
For example, the model proposed by \citet{2022MNRAS.516.4097P} effectively synthesized molecules CH$_3$OH, CH$_3$CHO, and CH$_3$OCH$_3$, but the abundance of the molecule HCOOCH$_3$ was two orders of magnitude lower than the observed value.
Even with adjustments to certain model parameters, achieving consistency between these COMs molecules and the observations remains unattainable.
The simulated results for these COMs were found to be lower by two to nine orders of magnitude compared to the observed values in both O-rich (C/O $\le$ 1) and C-rich (C/O $>$ 1)conditions in the work by \cite{2024AA...682A.109M}.
However, our multi-phase models, incorporating both radiolysis and sputtering mechanisms, effectively synthesized these COMs and demonstrated agreement with the observations.
The simulation results for COMs in our models, MC1 to MC4, all deviated from the observed values by less than a factor of 3.
Due to varying sputtering rates, the time corresponding to the coincidence of MC1 to MC4 with the observed COMs abundance also varied.
The MC1 model approximately matches the observed values at around 4.95$\times 10^5$ years, while the MC2 model aligns with observations at 2.93$\times 10^5$ years, the MC3 model at 3.94$\times 10^5$ years, and the MC4 model at 6.57$\times 10^5$ years.
Due to the absence of corresponding mechanisms, our MC5 and MC6 models could only synthesize CH$_3$OH.
Except for \cite{2022MNRAS.516.4627C}, who employed the Monte Carlo method, other studies utilized the rate equation. Although the Monte Carlo method requires more computational time, it effectively avoids the finite size effect when handling surface chemistry processes on dust grains, making it more suitable for our multiphase model and the newly introduced mechanisms.

\begin{deluxetable*}{llllllll}
\tablecaption{The abundance ratios of O-bearing COMs relative to H$_2$ (n$[\text{X}]$/n$_{\text{H}_2}$) in the different models, along with a comparison to observed species in TMC-1 and previous work.}
\tabletypesize{\scriptsize}
\label{tab:obesfit}
\tablehead{\colhead{Model} &\colhead{Time (yr)} & \colhead{CH$_3$OH}& \colhead{CH$_3$CHO}&
\colhead{HCOOCH$_3$}&\colhead{CH$_3$OCH$_3$} &\colhead{C$_2$H$_5$OH} &\colhead{Method}
}
\startdata
	Observations &-&(4.8$\pm$0.3)(-09)$^b$ & (3.5$\pm$0.2)(-10) $^b$&(1.1$\pm$0.2)(-10)$^c$&(2.5$\pm$0.7)(-10)$^c$&1.1$\pm$0.3(-10)$^d$\\
	MC1      & 4.95(5)& 4.15(-09) &1.22(-10) & 9.00(-11) & 3.80(-10) &2.62(-10)&MC-M\\
	MC2  & 2.93(5) & 2.50(-09)&1.94(-10)&5.20(-11)&2.64(-10)&1.86(-10)&MC-M\\
	MC3 & 3.94(5) &4.26(-09)&2.98(-10)&1.04(-10)&5.64(-10)&4.46(-10)&MC-M\\
	MC4 & 6.57(5) &4.34(-09)&9.40(-11)&4.60(-11)&3.08(-10)&1.98(-10)&MC-M\\
	MC5  & 4.65(5) &3.12(-09)&{\bf 1.20(-11)}&{\bf 2.00(-12)}&{\bf 0.00}&{\bf 8.00(-12)}&MC-M\\
	MC6 & 3.64(5)&2.01(-09) &{\bf 1.8(-11)}&{\bf 0.00}&{\bf 0.00}&{\bf 8.00(-12)}&MC-M\\
	PW1$^e$&-&{\bf 1.5(-10)}&3.2(-10)&-&6.6(-11)&-&NAUTILUS-3\\
	PW2$^f$&5.0(5)&3.5(-10)&2.0(-10)&{\bf 8.5(-12)}&3.7(-10)&-&NAUTILUS-3\\
	PW3$^g$  &6.0(5)&{\bf 1.0(-10)}& {\bf 2.0(-12)}&{\bf 5.0(-16)}&7.0(-11)&-&NAUTILUS-3\\
	PW4$^i$  &2.0(5)&5.40(-09)& 8.80(-11)& 1.14(-11)&{\bf 1.20(-11)}&-&MC-2\\
	PW5$^j$  &5.3(5)&5.29(-09)& {\bf 5.35(-12)}& {\bf 2.68(-12)}&{\bf 9.55(-13)}&-&NAUTILUS-2\\
	PW6$^h$  &1.6(5)&{\bf1.7(-15)}& {\bf 2.1(-12)}& {\bf 1.5(-16)}&{\bf 5.5(-15)}&{\bf 4.0(-14)}&only gas phase\\
	PW7$^k$  &1.0(6)&{\bf2.7(-13)}& {\bf 5.1(-12)}& {\bf 8.0(-19)}&{\bf 9.2(-19)}&{\bf 7.0(-13)}&only gas phase\\
\enddata
\tablecomments{\\
{\textup{a}}({\textup{b}}) = {\textup{a}} $\times$ 10$^{\textup{b}}$. \\
Bold font indicates underestimation by more than one order of magnitude compared with the observed values.\\
$^b$Observational data are taken from \citet{2020AA...642L..17C}.\\
$^c$Observational data are taken from \citet{2021AA...649L...4A}.\\
$^d$Observational data are taken from \citet{2023AA...673A..34A}.\\
$^e$The data comes from the `Updated model’ presented in \citet{2023MNRAS.519.4622C} paper.\\
$^f$The data comes from the `6HSM model’ presented in \citet{2022MNRAS.516.4097P} paper.\\
$^g$The data comes from the `Fig. 4. Sputtering model’ presented in \citet{2021AA...652A..63W} paper.\\
$^i$The data comes from the `ref. model’ presented in \citep{2022MNRAS.516.4627C} paper.\\
$^j$The data comes from the `best-fit model’ presented in \citep{2022ApJ...928..175C} paper.\\
$^h$ The data comes from the `best-fit’ values presented in \citep{2024AA...682A.109M} paper. O-rich conditions: C/O=0.75.\\
$^k$ The data comes from the `best-fit’ values presented in \citep{2024AA...682A.109M} paper. C-rich conditions: C/O=1.4.\\
MC-M: This work. Monte Carlo Multi-phase Model.\\
MC-2: Monte Carlo. Two-phase Model.\\
NAUTILUS-3: Rate Equation. NAUTILUS three-phase package.\\
NAUTILUS-2: Rate Equation. NAUTILUS two-phase package.\\
}
\end{deluxetable*}

Table \ref{obsTMC1} shows the abundance ratios of species relative to H$_2$ at the global best-fitting time for different models, along with a comparison to observed species abundance in TMC-1. Bold font signals over or under one order of magnitude compared to observed values.
When comparing the simulation results of 94 molecules with observations, some simulated species abundances are zero due to the use of the Monte Carlo method. As a result, calculating D(t) using Equation \ref{eq_best_local} yields values approaching positive infinity. To address this,
we defined the global best-fitting time as the moment when, among the 94 species, the number of simulated species abundances and observed values within one order of magnitude was maximized.
 Appendix \ref{appendix A} also provides the method and results of calculating the global best-fitting time using Equation \ref{eq_best_local}.
Since different observations can yield varying results for the same molecule, such as \citet{2021AA...648L...3C} reporting an H$_2$CS abundance of 4.7$\times10^{-9}$ and \citet{2013ChRv..113.8710A} reporting 7.0$\times10^{-10}$, we have prioritized using the most recent observational data wherever possible.
In our comparative analysis of simulation values for 94 species with observed data, we investigated of global best-fitting time across various models. 
The MC1 model showed its optimal fit at 2.42$\times10^5$ years, with 61 species falling within one order of magnitude of the observed values, accounting for approximately 64.89\% of the species (the fraction of reproduced species values). Similarly, the MC2 and MC3 models each had 63 species closely matching the observed values, representing approximately 67.02\%. However, their global best-fitting times differ, with the MC2 model at 2.32$\times10^5$ years and the MC3 model at 2.63$\times10^5$ years. The MC4 model achieved its optimal fit at 1.72$\times 10^5$ years, with 58 species within one order of magnitude of the observed values, representing around 61.70\%. In comparison, the MC5 model's optimal time was 1.31$\times10^5$ years, with 57 species closely matching the observations, accounting for approximately 60.64\%. Lastly, the MC6 model reached its optimal time at 6.06$\times10^4$ years, with 58 species within one order of magnitude of the observed values, representing about 61.70\%.
MC2 and MC3 stands out as the model with the best fit, followed by MC1 and MC4. 
MC5 and MC6, lacking relevant mechanisms, exhibit approximately 6\% lower consistency compared to the MC2 and MC3 models.
Our models, incorporating radiolysis and sputtering mechanisms effectively synthesized O-bearing COMs at the global best-fitting time and were in agreement with observations within one order of magnitude.

\startlongtable
\clearpage
\begin{deluxetable*}{llllllll}
\tablecaption{The abundance ratios of species relative to H$_2$ at the global best-fitting times for different models, along with a comparison to observed species in TMC-1.}
\tabletypesize{\scriptsize}
\tablehead{
\colhead{Species} & \colhead{Observed} &\colhead{MC1} &
\colhead{MC2} & \colhead{MC3} & \colhead{MC4} & \colhead{MC5} & \colhead{MC6}
}
\startdata
C$_2$					    &5.0(-08) $^1$  &{\bf 5.24(-10)}  &{\bf 6.02(-10)}  &{\bf 4.82(-10)}  &{\bf 1.12(-9)}  &{\bf 2.03(-9)}  &{\bf 3.58(-9)}  \\
CH						    &2.0(-08) $^2$  &2.54(-9)  &3.03(-9)  &2.41(-9)  &4.48(-9)  &5.84(-9)  &7.97(-9)  \\
CO						    &1.7(-04) $^2$  &1.33(-4)  &1.35(-4)  &1.33(-4)  &1.29(-4)  &1.17(-4)  &8.33(-5)  \\
CS					  	    &3.5(-08) $^3$  &{\bf 1.21(-9)}  &{\bf 1.41(-9)}  &{\bf 1.07(-9)}  &{\bf 1.59(-9)}  &{\bf 1.88(-9)}  &{\bf 3.20(-9)}  \\
CN					  	    &5.0(-09)$^4$  &1.00(-8)  &1.32(-8)  &8.94(-9)  &1.87(-8)  &2.47(-8)  &2.98(-8)  \\
OH					  	    &3.0(-07)$^2$  &5.47(-8)  &5.35(-8)  &5.97(-8)  &3.56(-8)  &{\bf 2.51(-8)}  &{\bf 1.61(-8)}  \\
O$_2$					    &$<$3.0(-06) $^{5}$  &6.04(-7)  &3.91(-7)  &9.36(-7)  &9.66(-8)  &4.02(-8)  &1.57(-8)  \\
NO				  		    &2.7(-08) $^2$  &1.06(-8)  &1.10(-8)  &1.29(-8)  &5.60(-9)  &3.23(-9)  &{\bf 1.68(-9)}  \\
NS 				  		    &1.7(-10) $^6$  &{\bf 0.00(0)}  &{\bf 2.00(-12)}  &{\bf 2.00(-12)}  &{\bf 2.00(-12)}  &{\bf 0.00(0)}  &{\bf 2.00(-12)}  \\
NS$^+$ 				  	    &5.2(-12) $^6$  &2.00(-12)  &4.00(-12)  &2.00(-12)  &{\bf 0.00(0)}  &{\bf 0.00(0)}  &2.00(-12)  \\
SO			  			    &1.5(-09) $^2$  &1.32(-9)  &8.56(-10)  &1.79(-9)  &2.04(-10)  &{\bf 8.80(-11)}  &{\bf 5.20(-11)}  \\
C$_2$H				  	    &6.5(-08) $^7$  &{\bf 2.36(-9)}  &{\bf 2.56(-9)}  &{\bf 2.32(-9)}  &{\bf 2.49(-9)}  &{\bf 2.67(-9)}  &{\bf 2.98(-9)}  \\
C$_2$O			  		    &7.5(-11) $^9$  &{\bf 2.00(-12)}  &{\bf 2.00(-12)}  &{\bf 0.00(0)}  &{\bf 0.00(0)}  &8.00(-12)  &1.20(-11)  \\
C$_2$S			  		    &7.0(-10) $^8$  &1.32(-9)  &1.23(-9)  &1.32(-9)  &1.29(-9)  &1.21(-9)  &8.26(-10)  \\
HCS						    &5.5(-10) $^3$  &{\bf 0.00(0)}  &{\bf 0.00(0)}  &{\bf 2.00(-12)}  &{\bf 2.00(-12)}  &{\bf 0.00(0)}  &{\bf 0.00(0)}  \\
HCO					  	    &1.1(-10)$^9$  &4.20(-11)  &3.20(-11)  &5.00(-11)  &6.00(-11)  &8.60(-11)  &1.04(-10)  \\
HCN						    &1.1(-08) $^2$  &1.14(-8)  &2.13(-8)  &1.25(-8)  &3.79(-8)  &6.67(-8)  &9.27(-8)  \\
HNC						    &2.6(-08) $^2$  &1.17(-8)  &2.08(-8)  &1.19(-8)  &4.19(-8)  &7.36(-8)  &9.91(-8)  \\
H$_2$S					    &$<$5.0(-10)$^2$  &5.40(-11)  &4.40(-11)  &4.40(-11)  &1.40(-11)  &1.20(-11)  &1.60(-11)  \\
H$_2$O					    &$<$7.0(-08) $^2$  &{\bf 1.58(-6)}  &{\bf 1.88(-6)}  &{\bf 1.54(-6)}  &{\bf 2.38(-6)}  &{\bf 2.47(-6)}  &{\bf 1.58(-6)}  \\
OCS						    &$<$1.8(-09)$^{10}$  &4.48(-10)  &5.22(-10)  &4.66(-10)  &6.20(-10)  &6.06(-10)  &4.06(-10)  \\
SO$_2$					    &3.0(-10)$^2$  &4.60(-11)  &3.20(-11)  &1.06(-10)  &{\bf 0.00(0)}  &{\bf 0.00(0)}  &{\bf 0.00(0)}  \\
HCS$^+$					    &1.0(-09)$^8$  &{\bf 0.00(0)}  &{\bf 0.00(0)}  &{\bf 0.00(0)}  &{\bf 2.00(-12)}  &{\bf 2.00(-12)}  &{\bf 4.00(-12)}  \\
HCO$^{+}$				    &9.3(-09) $^2$  &4.38(-9)  &4.11(-9)  &4.45(-9)  &3.22(-9)  &2.48(-9)  &1.69(-9)  \\
N$_2$H$^{+}$			  	    &2.8(-10) $^2$  &3.08(-10)  &2.82(-10)  &3.30(-10)  &1.36(-10)  &8.80(-11)  &3.40(-11)  \\
C$_3$N					    &1.2(-09) $^{11}$  &2.06(-10)  &2.42(-10)  &1.80(-10)  &3.30(-10)  &3.04(-10)  &1.80(-10)  \\
C$_3$S					    &1.3(-09) $^8$  &4.84(-10)  &5.70(-10)  &5.22(-10)  &6.18(-10)  &4.86(-10)  &3.66(-10)  \\
C$_3$O					    &1.2(-10)$^9$  &{\bf 1.80(-9)}  &{\bf 2.18(-9)}  &{\bf 1.51(-9)}  &{\bf 2.28(-9)}  &{\bf 1.89(-9)}  &1.20(-9)  \\
C$_3$H 					    &3.0(-09) $^{10}$  &2.11(-9)  &2.34(-9)  &2.08(-9)  &2.12(-9)  &1.98(-9)  &1.61(-9)  \\
HNCO					    &1.3(-09) $^{12}$  &{\bf 5.40(-11)}  &1.56(-10)  &{\bf 1.00(-10)}  &1.44(-10)  &2.50(-10)  &4.32(-10)  \\
H$_2$CO					    &5.0(-08) $^{13}$  &2.18(-8)  &3.33(-8)  &2.60(-8)  &4.23(-8)  &5.50(-8)  &5.36(-8)  \\
H$_2$CS					    &4.7(-09) $^3$  &{\bf 3.92(-10)}  &{\bf 3.98(-10)}  &{\bf 3.32(-10)}  &{\bf 4.32(-10)}  &{\bf 4.16(-10)}  &{\bf 2.40(-10)}  \\
H$_2$CN					    &1.5(-11) $^2$  &1.80(-11)  &3.20(-11)  &2.60(-11)  &3.40(-11)  &3.60(-11)  &3.00(-11)  \\
HCCN					    &(4.4$\pm$0.4)(-11)$^{14}$  &2.60(-11)  &7.20(-11)  &5.80(-11)  &{\bf 2.00(-12)}  &{\bf 0.00(0)}  &{\bf 0.00(0)}  \\
HC$_2$O 				    &7.7(-11) $^9$  &1.40(-11)  &2.80(-11)  &3.20(-11)  &{\bf 0.00(0)}  &{\bf 0.00(0)}  &{\bf 0.00(0)}  \\
NH$_3$					    &5.0(-08) $^{10}$  &3.12(-8)  &3.85(-8)  &4.07(-8)  &2.47(-8)  &2.09(-8)  &1.45(-8)  \\
C$_3$H$^+$				    &(2.4$\pm$0.2)(-12)$^{15}$  &8.00(-12)  &6.00(-12)  &8.00(-12)  &6.00(-12)  &1.60(-11)  &2.00(-11)  \\
HOCO$^{+}$ 				    &4.0(-11) $^{12}$  &6.00(-12)  &4.00(-12)  &1.00(-11)  &6.00(-12)  &{\bf 0.00(0)}  &{\bf 0.00(0)}  \\
HC$_2$S$^+$				    &$\leq$9.0(-11)$^8$  &2.00(-12)  &6.00(-12)  &2.00(-12)  &8.00(-12)  &2.00(-12)  &0.00(0)  \\
HCNH$^{+}$				    &1.9(-09) $^2$  &2.34(-10)  &4.40(-10)  &2.72(-10)  &7.12(-10)  &1.00(-9)  &1.31(-9)  \\
CH$_3$O					    &$<$1.0(-10) $^2$  &{\bf 4.10(-10)}  &{\bf 1.35(-9)}  &{\bf 1.61(-9)}  &8.40(-11)  &0.00(0)  &0.00(0)  \\
C$_4$N  					    &$\leq$4.0(-09)$^{14}$  &7.34(-10)  &8.34(-10)  &5.36(-10)  &1.11(-9)  &9.36(-10)  &6.38(-10)  \\
C$_4$H					    &8.5(-09) $^{11}$  &4.37(-9)  &5.64(-9)  &3.18(-9)  &6.73(-9)  &5.55(-9)  &3.75(-9)  \\
C$_3$H$_2$				    &1.24(-08)$^{16}$  &2.29(-8)  &1.90(-8)  &2.73(-8)  &8.25(-9)  &4.31(-9)  &1.71(-9)  \\
C$_4$S					    &(3.8$\pm$0.4)(-12)$^3$  &2.80(-11)  &3.80(-11)  &1.00(-11)  &{\bf 9.60(-11)}  &{\bf 1.14(-10)}  &{\bf 8.60(-11)}  \\
CH$_2$NH				    &$<$3.6(-0.9)$^1$  &1.50(-9)  &{\bf 3.81(-9)}  &2.67(-9)  &1.93(-9)  &1.45(-9)  &6.14(-10)  \\
HC$_3$O					    &1.3(-11)$^9$  &2.00(-12)  &6.00(-12)  &4.00(-12)  &{\bf 0.00(0)}  &{\bf 0.00(0)}  &{\bf 0.00(0)}  \\
HNC$_3$					    &(5.2$\pm$0.3)(-11) $^{17}$  &2.38(-10)  &2.52(-10)  &2.08(-10)  &2.18(-10)  &1.90(-10)  &1.06(-10)  \\
H$_2$CCO				    &1.4(-09)$^{12}$  &7.21(-9)  &1.10(-8)  &6.35(-9)  &{\bf 2.78(-8)}  &{\bf 4.55(-8)}  &{\bf 4.91(-8)}  \\
HC$_3$N					    &2.3(-08) $^{17}$  &{\bf 2.30(-9)}  &3.32(-9)  &2.38(-9)  &5.70(-9)  &9.14(-9)  &1.01(-8)  \\
HCOOH					    &1.4(-10)$^{12}$  &{\bf 1.92(-9)}  &{\bf 2.88(-9)}  &{\bf 2.77(-9)}  &{\bf 1.81(-9)}  &1.05(-9)  &2.68(-10)  \\
HC$_2$NC				    &(3.0$\pm$0.3)(-10) $^{17}$  &2.94(-10)  &4.40(-10)  &2.48(-10)  &6.66(-10)  &9.16(-10)  &8.10(-10)  \\
H$_2$CCN				    &1.47(-09)$^{18}$  &4.68(-10)  &5.90(-10)  &5.96(-10)  &5.58(-10)  &6.32(-10)  &4.74(-10)  \\
HC$_3$O$^{+}$			    &2.1(-11) $^{12}$  &2.20(-11)  &6.00(-12)  &1.00(-11)  &2.00(-11)  &8.00(-12)  &6.00(-12)  \\
H$_3$CO$^+$				    &$<$3.0(-11)$^{12}$  &{\bf 5.00(-11)}  &{\bf 7.00(-11)}  &{\bf 6.00(-11)}  &{\bf 7.60(-11)}  &{\bf 8.00(-11)}  &{\bf 7.00(-11)}  \\
HC$_3$S$^+$				    &(2.0$\pm$0.5)(-11)$^8$  &6.00(-12)  &4.00(-12)  &6.00(-12)  &4.00(-12)  &4.00(-12)  &6.00(-12)  \\
C$_5$H					    &1.3(-10) $^{19}$  &2.76(-10)  &4.16(-10)  &2.30(-10)  &4.94(-10)  &4.90(-10)  &3.88(-10)  \\
C$_5$N					    &4.7(-11) $^{11}$  &1.80(-11)  &3.20(-11)  &2.00(-11)  &6.60(-11)  &4.80(-11)  &5.20(-11)  \\
CH$_3$OH				    &(4.8)(-09)$^{12}$  &1.06(-9)  &1.55(-9)  &1.80(-9)  &{\bf 2.40(-10)}  &{\bf 2.20(-11)}  &{\bf 4.20(-11)}  \\
CH$_3$CN				    &4.7(-10) $^{18}$  &1.92(-10)  &6.04(-10)  &1.82(-10)  &1.92(-9)  &4.42(-9)  &4.41(-9)  \\
CH$_2$CCH				    &8.7(-09)$^{20}$  &{\bf 6.60(-11)}  &{\bf 9.40(-11)}  &{\bf 9.60(-11)}  &{\bf 6.20(-11)}  &{\bf 6.60(-11)}  &{\bf 2.40(-11)}  \\
H$_2$C$_4$				    &3.3(-10) $^{21}$  &5.26(-10)  &4.70(-10)  &5.36(-10)  &2.32(-10)  &1.22(-10)  &4.60(-11)  \\
NH$_2$CHO				    &$\leq$5.0(-12)$^{12}$  &{\bf 8.00(-12)}  &{\bf 4.20(-11)}  &{\bf 5.00(-11)}  &0.00(0)  &2.00(-12)  &0.00(0)  \\
H$_2$CCCN				    &(2.5$\pm$0.4)(-11)$^{22}$  &{\bf 0.00(0)}  &{\bf 0.00(0)}  &{\bf 0.00(0)}  &{\bf 0.00(0)}  &{\bf 0.00(0)}  &{\bf 0.00(0)}  \\
HCCCHO					    &(1.5$\pm$0.3)(-10)$^{13}$  &3.00(-11)  &2.80(-11)  &2.60(-11)  &2.40(-11)  &2.00(-11)  &3.00(-11)  \\
C$_5$H$^+$				    &(8.8$\pm$0.5)(-12)$^{15}$  &{\bf 0.00(0)}  &2.00(-12)  &{\bf 0.00(0)}  &3.00(-11)  &7.80(-11)  &{\bf 1.26(-10)}  \\
CH$_3$CO$^+$			    &(3.2$\pm$0.3)(-11)$^{23}$  &2.20(-11)  &3.80(-11)  &2.20(-11)  &4.80(-11)  &6.00(-11)  &7.40(-11)  \\
H$_2$CCCH$^+$			    &(7.0$\pm$1.5)(-11)$^{24}$  &2.98(-10)  &3.24(-10)  &2.78(-10)  &3.62(-10)  &2.92(-10)  &2.28(-10)  \\
HCCNCH$^+$				    &(3.0$\pm$0.5)(-12)$^{25}$  &2.00(-12)  &{\bf 0.00(0)}  &2.00(-12)  &2.00(-12)  &1.00(-11)  &4.00(-12)  \\
HC$_3$NH$^+$			    &1.0(-10)$^2$  &5.60(-11)  &7.00(-11)  &4.20(-11)  &1.12(-10)  &1.62(-10)  &1.78(-10)  \\
C$_6$H					    &4.8(-10) $^{11}$  &1.06(-10)  &9.00(-11)  &8.20(-11)  &9.20(-11)  &1.08(-10)  &1.54(-10)  \\
CH$_3$CHO				    &(3.5$\pm0.2$)(-10)$^{12}$  &{\bf 3.20(-11)}  &9.60(-11)  &8.40(-11)  &{\bf 1.00(-11)}  &{\bf 2.00(-12)}  &{\bf 0.00(0)}  \\
CH$_3$CCH				    &1.2(-08)$^{10}$  &{\bf 6.00(-12)}  &{\bf 4.00(-12)}  &{\bf 4.00(-12)}  &{\bf 0.00(0)}  &{\bf 0.00(0)}  &{\bf 0.00(0)}  \\
CH$_2$CHCN				    &(6.5$\pm$0.5)(-10)$^{14}$  &{\bf 2.00(-12)}  &{\bf 0.00(0)}  &{\bf 0.00(0)}  &{\bf 6.00(-12)}  &{\bf 0.00(0)}  &{\bf 0.00(0)}  \\
l-H$_2$C$_5$			  	    &(1.8$\pm$0.5)(-12)$^{21}$  &{\bf 1.04(-10)}  &{\bf 5.20(-11)}  &{\bf 1.00(-10)}  &{\bf 2.80(-11)}  &8.00(-12)  &2.00(-12)  \\
HC$_5$N					    &1.8(-08) $^{17}$  &{\bf 8.80(-11)}  &{\bf 1.74(-10)}  &{\bf 6.00(-11)}  &{\bf 9.48(-10)}  &2.21(-9)  &3.01(-9)  \\
CH$_2$CHCCH			    &(1.2$\pm$0.2)(-09)$^{14}$  &{\bf 2.00(-12)}  &{\bf 2.00(-12)}  &{\bf 2.00(-12)}  &{\bf 0.00(0)}  &{\bf 2.00(-12)}  &{\bf 0.00(0)}  \\
C$_7$H					    &6.5(-12)$^7$  &5.00(-11)  &5.60(-11)  &6.40(-11)  &4.20(-11)  &4.00(-11)  &1.20(-11)  \\
CH$_3$C$_3$N			    &(8.66$\pm$0.46)(-11) $^{26}$  &{\bf 4.00(-12)}  &2.00(-11)  &1.00(-11)  &6.20(-11)  &9.40(-11)  &7.00(-11)  \\
H$_2$C$_6$				    &8.0(-12) $^{21}$  &4.80(-11)  &4.80(-11)  &{\bf 9.80(-11)}  &1.60(-11)  &4.00(-12)  &4.00(-12)  \\
HCOOCH$_3$				    &(1.1$\pm$0.2)(-10)$^{27}$  &1.60(-11)  &2.20(-11)  &1.20(-11)  &{\bf 0.00(0)}  &{\bf 0.00(0)}  &{\bf 0.00(0)}  \\
HC$_5$NH$^+$			    &(7.5$\pm$2.2)(-11)$^{28}$  &{\bf 0.00(0)}  &{\bf 0.00(0)}  &{\bf 0.00(0)}  &{\bf 0.00(0)}  &{\bf 0.00(0)}  &{\bf 4.00(-12)}  \\
C$_8$H					    &4.6(-11) $^1$  &1.60(-11)  &8.00(-12)  &1.40(-11)  &6.00(-12)  &{\bf 4.00(-12)}  &8.00(-12)  \\
C$_2$H$_5$OH			    &(1.1$\pm$0.3)(-10)$^{29}$  &3.40(-11)  &8.00(-11)  &9.60(-11)  &1.40(-11)  &{\bf 0.00(0)}  &{\bf 0.00(0)}  \\
CH$_3$OCH$_3$			    &(3.5$\pm$0.7)(-10)$^{27}$  &4.00(-11)  &1.02(-10)  &1.14(-10)  &{\bf 2.00(-12)}  &{\bf 0.00(0)}  &{\bf 0.00(0)}  \\
CH$_3$C$_4$H			    &(1.0$\pm$0.06)(-09)$^{26}$  &{\bf 2.80(-11)}  &{\bf 2.60(-11)}  &{\bf 3.00(-11)}  &{\bf 2.20(-11)}  &{\bf 4.00(-12)}  &{\bf 2.00(-12)}  \\
HC$_7$N					    &(2.1$\pm$0.2)(-09)$^{30}$  &{\bf 8.00(-12)}  &{\bf 1.80(-11)}  &{\bf 2.20(-11)}  &{\bf 5.60(-11)}  &{\bf 1.08(-10)}  &2.16(-10)  \\
C$_6$H$_4$					   &(5.0)(-11)$^{32}$  &6.00(-12)  &8.00(-12)  &1.20(-11)  &8.00(-12)  &{\bf 4.00(-12)}  &{\bf 0.00(0)}  \\
C$_9$H					   &$\leq$3.5(-12)$^{7}$  &0.00(0)  &2.00(-12)  &2.00(-12)  &2.00(-12)  &2.00(-12)  &2.00(-12)  \\
CH$_3$C$_5$N			    &(9.5$\pm$0.9)(-12) $^{31}$  &2.00(-12)  &{\bf 0.00(0)}  &2.00(-12)  &1.40(-11)  &2.20(-11)  &1.60(-11)  \\
HC$_7$NH$^+$			    &(5.5$\pm$0.7)(-12)$^{30}$  &{\bf 0.00(0)}  &{\bf 0.00(0)}  &{\bf 0.00(0)}  &2.00(-12)  &2.00(-12)  &{\bf 0.00(0)}  \\
HC$_9$N					    &1.1(-09)$^{10}$  &{\bf 0.00(0)}  &{\bf 4.00(-12)}  &{\bf 8.00(-12)}  &{\bf 1.80(-11)}  &{\bf 2.40(-11)}  &{\bf 1.40(-11)}  \\
CH$_3$C$_6$H			    &(7.0$\pm$0.7)(-11)$^{31}$  &{\bf 0.00(0)}  &{\bf 0.00(0)}  &{\bf 0.00(0)}  &{\bf 0.00(0)}  &{\bf 0.00(0)}  &{\bf 0.00(0)}  \\
CH$_3$C$_7$N			    &(8.6$\pm$0.19)(-12)$^{26}$  &{\bf 0.00(0)}  &{\bf 0.00(0)}  &{\bf 0.00(0)}  &{\bf 0.00(0)}  &2.00(-12)  &2.00(-12)  \\
\hline
Total& & 61&63&63&58&57&58\\
global best-fitting time(yr)& & 2.42$\times 10^{5}$&2.32$\times 10^{5}$&2.63$\times 10^{5}$&1.72$\times 10^{5}$&1.31$\times 10^{5}$&6.06$\times 10^{4}$\\
\enddata
\tablecomments{Adopting a column density of H$_2$ of 10$^{22}$ cm$^{-2}$ for TMC-1 \citep{1987AA...176..299C}.\\
{\textup{a}}({\textup{b}}) = {\textup{a}} $\times$ 10$^{\textup{b}}$. \\
Bold font signals over or under one order of magnitude compared to observed values.\\
$^1$Observational data are taken from \citet{2013AA...550A..36M}, $^2$Observational data are taken from \citet{2013ChRv..113.8710A}.\\
$^3$Observational data are taken from \citet{2021AA...648L...3C}, $^4$Observational data are taken from \citet{2009ApJ...700..752W}.\\
$^5$Observational data are taken from \citet{2002AA...395..233R}, $^6$Observational data are taken from \citet{2018ApJ...853L..22C}.\\
$^7$Observational data are taken from \citet{2022AA...663L...9C}, $^8$Observational data are taken from \citet{2021AA...646L...3C}.\\
$^9$Observational data are taken from \citet{2021AA...656L..21C}, $^{10}$Observational data are taken from \citet{2016ApJS..225...25G}.\\
$^{11}$Observational data are taken from \citet{2023AA...677A.106A}, $^{12}$Observational data are taken from \citet{2020AA...642L..17C}. \\
$^{13}$Observational data are taken from \citet{2021AA...650L..14C}, $^{14}$Observational data are taken from \citet{2021AA...647L...2C}.\\
$^{15}$Observational data are taken from \citet{2022AA...657L..16C}, $^{16}$Observational data are taken from \citet{2017MNRAS.470.4075L}.\\
$^{17}$Observational data are taken from \citet{2020AA...642L...8C}, $^{18}$Observational data are taken from \citet{2021AA...646L...1C}. \\
$^{19}$Observational data are taken from \citet{2022AA...663L...2C}, $^{20}$Observational data are taken from \citet{2021AA...647L..10A}. \\
$^{21}$Observational data are taken from \citet{2021AA...650L...9C}, $^{22}$Observational data are taken from \citet{cabezas2023laboratory}.\\
$^{23}$Observational data are taken from \citet{2021AA...646L...7C}, $^{24}$Observational data are taken from \citet{2023AA...676L...1S}.\\
$^{25}$Observational data are taken from  \citet{2022AA...659L...9A}, $^{26}$Observational data are taken from \citet{2022ApJ...924...21S}.\\
$^{27}$Observational data are taken from  \citet{2021AA...649L...4A}, $^{28}$Observational data are taken from \citet{2020AA...643L...6M}.\\
$^{29}$Observational data are taken from  \citet{2023AA...673A..34A}, $^{30}$Observational data are taken from \citet{2022AA...659L...8C}.\\
$^{31}$Observational data are taken from  \citet{2022AA...663L...3F}, $^{32}$Observational data are taken from \citet{2021AA...652L...9C}.\\}
\label{obsTMC1}
\end{deluxetable*}

\begin{figure*}
	\plotone{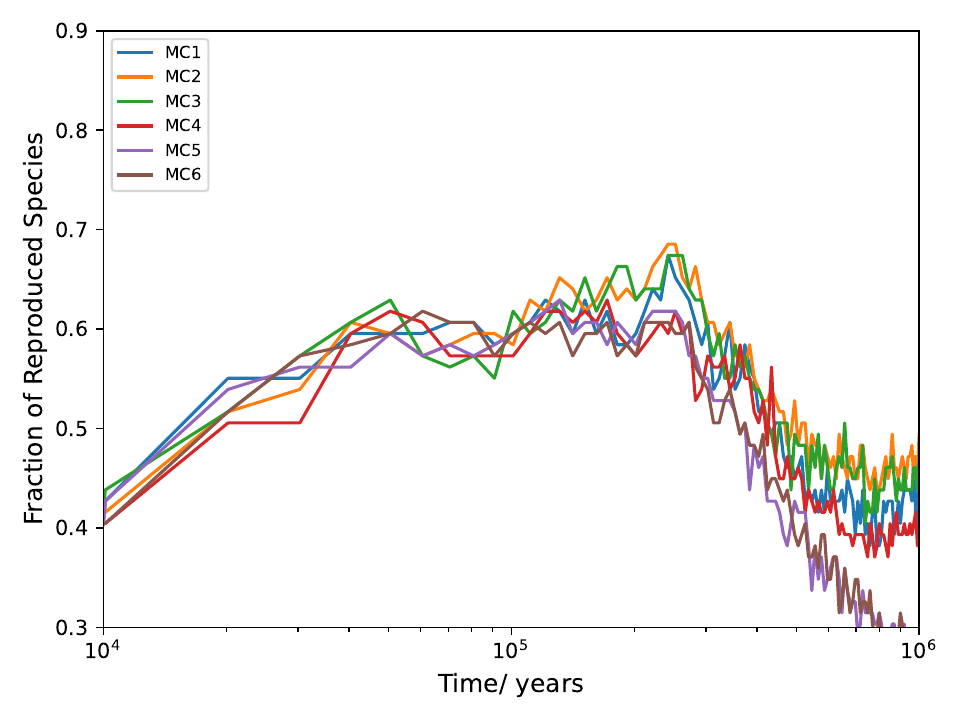}
    \caption{The Fraction of Reproduced Species Over Time in Different Models.}
    \label{fig:Fractionofperc}
\end{figure*}

Figure \ref{fig:Fractionofperc} displays the fraction of reproduced species over time for 94 species across different models.
The trend in the fraction of reproduced species was consistent across all models, showing an increase with time evolution, reaching a peak, and ultimately decreasing with further time evolution. 
In the MC1 to MC4 models, the fraction of reproduced species peaks around 2$\times10^5$ years and then begins to decline, eventually stabilizing. For the MC2 and MC3 models, this fraction stabilizes at approximately 45\%, while for the MC1 and MC4 models, it is slightly lower, around 40\%. In contrast, the MC5 and MC6 models gradually decline after reaching their peak and finally stabilize at around 25\%.
This difference is due to the inclusion of new mechanisms in the models, which allow species to desorb from dust ice mantles into the gas phase at later stages, where they continue to participate in further chemical reactions.

\subsection{Impact of C/O Ratios on the Results}
\begin{figure*}
	\plotone{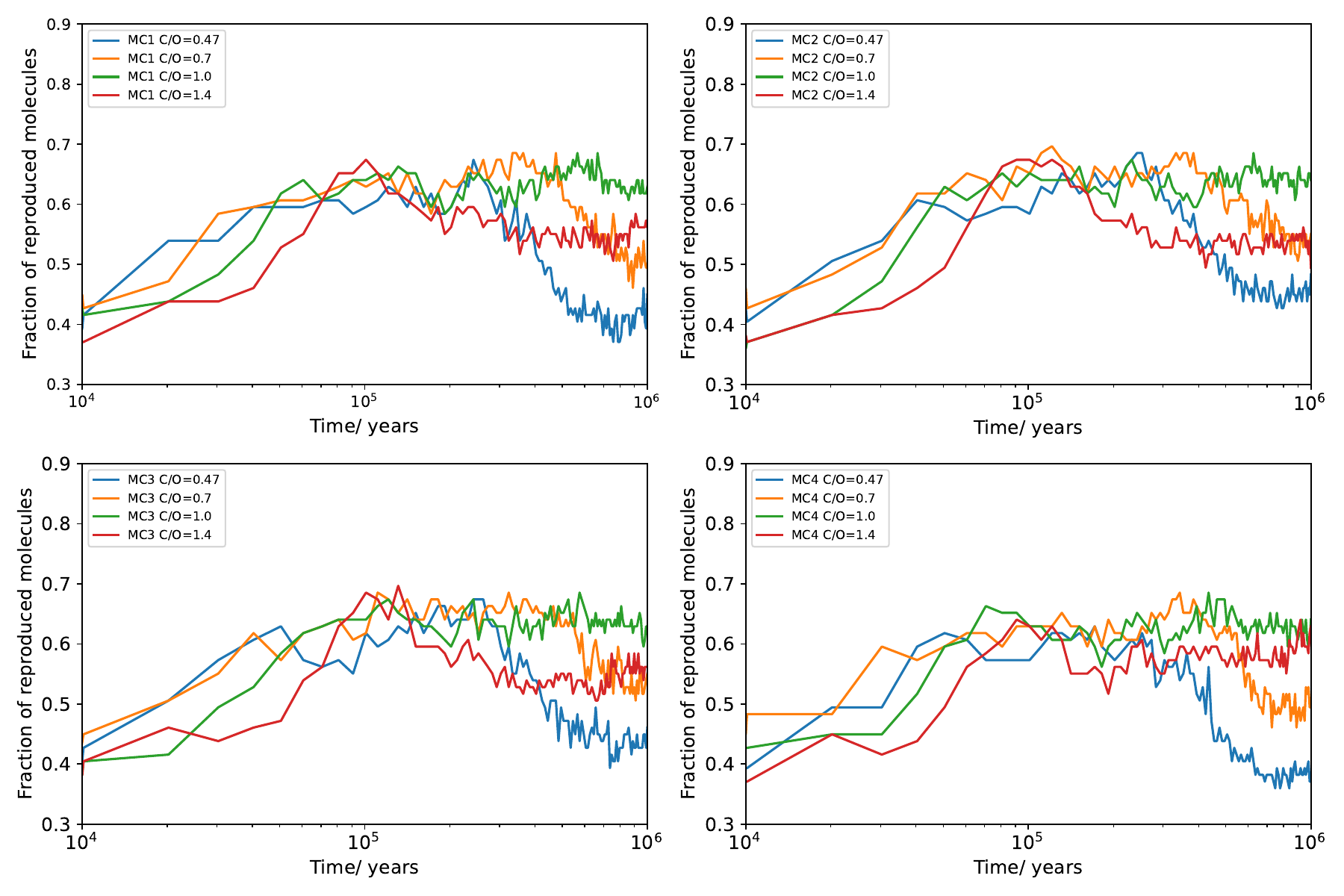}
    \caption{The fraction of reproduced species over time in different C/O ratios and models.}
    \label{fig:Fractionofpercc/o}
\end{figure*}

To investigate the influence of the C/O values in our model on the chemical evolution of TMC-1, we conducted relevant studies using the MC1 to MC4 models, each featuring one of four distinct C/O values: 0.47, 0.7, 1.0 and 1.4.
Figure \ref{fig:Fractionofpercc/o} shows the temporal evolution of the fraction of reproduced species values in the MC1 to MC4 models for varying C/O ratios. 
In our four models, the results showed that the peak values of the fraction of reproduced species did not vary significantly with different C/O ratios.

In the MC1 model, the peak values of the fraction of reproduced species remain nearly identical across different C/O ratios, at approximately 70\%. However, the times at which these peak values are reached differ. When C/O = 1.4, the fraction of reproduced species reaches its peak the earliest, around 1$\times 10^5$ years. When C/O $\leq$ 1.0, the time to reach the peak value is delayed as the C/O ratio increases. The peak times for C/O ratios of 0.47, 0.7, and 1.0 are approximately 2.5$\times10^5$, 4.5$\times10^5$, and 6.8$\times10^5$ years, respectively. At the final evolution time point of 1$\times10^6$ years, the fraction of reproduced species for different C/O values stabilizes with minor fluctuations. For C/O=1.0, the fraction is the highest at around 63\%, followed by C/O=1.4 at 57\%, C/O=0.7 at 54\%, and C/O=0.47 at 45\%.
The results for the other three models are consistent with those of the MC1 model. Changing the C/O ratio does not significantly alter the peak values of the fraction of reproduced species, although the time required to reach these peaks may vary slightly.

Previous studies have shown that different C/O ratios can result in varying peak values of the fraction of reproduced species. Table \ref{tab:allmec} shows the C/O values used in different studies when investigating the observed species in TMC-1.

During the 2000s, researchers began investigating the impact of different C/O ratios on species formation in the TMC-1 molecular cloud. Studies by \citet{2002AA...395..233R}, \citet{2004MNRAS.350..323S}, and \citet{2024AA...682A.109M} indicated that, under similar conditions, increasing the C/O ratio (C/O $>$ 1) led to better agreement between theoretical simulations and observations. However, in recent years, some studies, such as \citet{2022ApJ...928..175C}, have reported contrasting results.
In our models, varying the C/O ratio has minimal impact on the theoretical simulation results. Carbon atoms and ions tend to form CO, which is relatively more stable, leading to carbon being locked in CO and less available for forming other species. Additionally, some carbon can adsorb onto dust grains. Thus, increasing the C/O ratio allows the excess carbon to participate in reactions, leading to the formation of other molecules.

In our models that include the new mechanisms, carbon-containing species are still initially formed through gas-phase reactions before carbon gets locked in CO. At later stages, the small carbon-bearing species can desorb from dust ice mantle into the gas phase and participate in further reactions, leading to the formation of related molecules. For instance, the abundances of C$_2$H$_5$, C, C$_2$, and C$_3$ are one to two orders of magnitude higher in our models with the new mechanisms than in those without. This is also why different C/O ratios have a minimal effect on our models.

In recent observations of TMC-1, numerous COMs like HC$_7$O and CH$_3$COCH$_3$, long carbon chain molecules such as HC$_{11}$N and CH$_3$C$_8$H, and polycyclic aromatic hydrocarbons (PAHs) like c-C$_9$H$_8$ and c-C$_{10}$H$_7$CN have been detected. These findings suggest chemical diversity in TMC-1, hinting at potential gaps in our understanding of the underlying mechanisms.

According to the research results of \citet{2024AA...682A.109M}, TMC-1 is more likely to be a carbon-rich (C/O=1.4) cold molecular cloud with an age of approximately 1.0$\times 10^6$ years. 
In contrast, \citet{2022ApJ...928..175C} suggests that TMC-1 is an oxygen-rich (C/O = 0.5) cold molecular cloud with an age of around 5.3$\times 10^5$ years. 
However, our models cannot determine a specific C/O value, as varying C/O values have minimal impact on the overall fitting results under the new mechanisms.
This discrepancy could arise from differences in astrochemical model, chemical reaction networks, reaction mechanisms, etc. Millar et al.'s chemical reaction network is more comprehensive than ours, but they only consider gas phase chemical model, whereas our chemical model is a multiphase model, and have incorporated some new mechanisms. Currently, we have no means to confirm the precise C/O ratio in TMC-1.

\begin{deluxetable*}{llll}
\tablecaption{Different C/O ratio results in TMC-1.}
\tabletypesize{\scriptsize}
\label{tab:allmec}
\tablehead{\colhead{work} &\colhead{Best fitting time (yr)} & \colhead{C/O}& \colhead{Species}
}
\startdata
		Roberts and Herbst$^1$&5$\times 10^6 \sim$ 1$\times 10^7$&{\bf 1.0}&all species\\
		Smith et al.$^2$&1$\times 10^5$&0.42, 0.8, {\bf 1.2}&all species\\
		Wakelam et al. $^3$&(3$\sim$8)$\times 10^4$&0.4, {\bf 1.2}&all species\\
		Walsh et al.$^4$&(1$\sim$2)$\times 10^5$&0.4&all species\\
		Ag\'{u}ndez and Wakelam$^5$&(2$\sim$3)$\times 10^5$ or $>10^6$&0.55, {\bf 1.4}&all species\\
		McElroy et al.$^6$&1.7$\times 10^5$&0.44&all species\\
		Loison et al.$^7$&$\sim10^5$ &0.7&specific species (COMs)\\
		Vidal et al.$^8$&1$\times 10^6$&0.7&all species\\
		Majumdar et al.$^9$&1$\times 10^5$ or 3$\times 10^5$&0.7&specific species (deuteration species) \\
		Loison et al.$^{10}$&-&0.7&specific species (Carbon Chain species)\\
		Choi et al.$^{11}$&-&0.4$\sim$0.5&all species\\
		Maffucci et al. $^{12}$&-&0.7 and {\bf 1.2}&all species\\
		Iqbal and Wakelam$^{13}$&2$\times 10^5$&0.7&all species\\
		Ge et al.$^{14}$&3.3$\times 10^5$ or 1.7$\times 10^5$&0.47& specific species (PAH)\\
		McGuire et al.$^{15}$&-&{\bf 1.1}&specific species (Carbon Chain species) \\
		Cabezas et al.$^{16}$&-&0.75&specific species (deuteration species)\\
		Cabezas et al.$^{17}$&-&0.75&specific species (deuteration species)\\
		Cernicharo et al.$^{18}$&-&0.55& specific species(sulfur-bearing species)\\
		Ag\'{u}ndez et al.$^{19}$&-&0.55&specific species (Carbon Chain species)\\
		Fuentetaja et al.$^{20}$&-&0.55&specific species (Carbon Chain species)\\
		Fuentetaja et al.$^{21}$&- &0.55 & specific species (Carbon Chain species)\\
		Paulive et al.$^{22}$&5$\times 10^5$&0.7& specific species (COMs)\\
		Cabezas et al.$^{23}$&2.0$\times 10^5$&0.44&specific species (Carbon Chain species)\\
		Chen et al.$^{24}$&5.3$\times 10^5$&0.5&all species\\
		Carder et al.$^{25}$&-&0.7&specific species (COMs)\\
		Ag\'{u}ndez et al.$^{26}$&2$\times 10^5$&0.7&specific species (COMs)\\
		Cooke et al$^{27}$&3.7$\times 10^5$&{\bf 1.1}&specific species (Carbon Chain species)\\
		Carder et al.$^{28}$&-&0.7&specific species (COMs)\\
		Mebel et al.$^{29}$&-&0.55&specific species (Carbon Chain species)\\
		Millar et al.$^{30}$ &1.6$\times 10^5$ &0.44 &all species\\
		Millar et al.$^{03}$ &1.0$\times 10^6$ &{\bf 1.4} &all species\\
		\hline
	\enddata
	\tablecomments{\\
specific species: study single or a few molecules.\\
all species: studying molecules observed in TMC-1.\\
PAH: polycyclic aromatic hydrocarbons.\\
$^{1}$ \citet{2002AA...395..233R}\\
$^{2}$ \citet{2004MNRAS.350..323S}\\
$^{3}$ \citet{2006AA...451..551W}\\
$^{4}$ \citet{2009ApJ...700..752W}\\
$^{5}$ \citet{2013ChRv..113.8710A}\\
$^{6}$ \cite{2013AA...550A..36M}\\
$^{7}$ \citet{2016MNRAS.456.4101L}\\
$^{8}$ \citet{2017MNRAS.469..435V}\\
$^{9}$ \citet{2017MNRAS.466.4470M}\\
$^{10}$ \citet{2017MNRAS.470.4075L}\\
$^{11}$ \citet{2017ApJS..229...38C}\\
$^{12}$ \citet{2018ApJ...868...41M}\\
$^{13}$ \citet{2018AA...615A..20I}\\
$^{14}$ \citet{2020MNRAS.497.3306G}\\
$^{15}$ \citet{2020ApJ...900L..10M}\\
$^{16}$ \citet{2021AA...646L...1C}\\
$^{17}$ \citet{2021AA...650L..15C}\\
$^{18}$ \citet{2021AA...648L...3C}\\
$^{19}$ \citet{2021AA...647L..10A}\\
$^{20}$ \citet{2022AA...667L...4F}\\
$^{21}$ \citet{2022AA...663L...3F}\\
$^{22}$ \citet{2022MNRAS.516.4097P}\\
$^{23}$ \citet{2022AA...659L...8C}\\
$^{24}$ \citet{2022ApJ...928..175C}\\
$^{25}$ \citet{2023MNRAS.519.4622C}\\
$^{26}$ \citet{2023AA...673A..34A}\\
$^{27}$ \citet{2023ApJ...948..133C}\\
$^{28}$ \citet{2023MNRAS.519.4622C}\\
$^{29}$ \citet{2023ApJ...945L..40M}\\
$^{30}$ \citet{2024AA...682A.109M}\\
}
\end{deluxetable*}

\section{Conclusions}
\label{Conclusions1}
This article primarily investigates the formation mechanisms of oxygen-obeying COMs within the cold cloud core TMC-1. We employ a multiphase astrochemical model developed by us, incorporating with cosmic ray-induced radiolysis and sputtering mechanisms. The major outcomes of our research are summarized as follows:
\begin{enumerate}
  \item Our multiphase models, which incorporate sputtering and radiolysis mechanisms, effectively synthesize O-bearing COMs under typical cold molecular cloud conditions.
  Simulated abundances of molecules such as CH$_3$OH, HCOOCH$_3$, CH$_3$OCH$_3$, CH$_3$CHO, and C$_2$H$_5$OH agree with observations, within a factor of 3. 
The new mechanism facilitates the efficient gas-phase formation of certain COMs, with their precursors predominantly forming within the ice mantle and subsequently desorbing into the gas phase through sputtering.
   \item Our model also successfully generated carbon chain species in agreement with observations, including hydrocarbon molecules, the cyanopolyyne family, the methylcyanopolyynes family, sulfur-bearing carbon chains, and more. After comparing the simulated and observed abundances of 94 species, 63 were within one order of magnitude from the observed, constituting 67.02\%.
    \item Based on previous studies, we expanded the excitation reaction network, as shown in Table \ref{Suprathermalreaction}. Excited molecules are not only used to form COMs but also have the same probability of reacting with nearby molecules. Our new model can alter the distribution of species between dust and gas phases, allowing some COMs to form via gas-phase reactions, such as CH$_3$CHO. The new model favors the late-stage formation of carbon-containing molecules like hydrocarbon family, cyanopolyyne family.
This primarily occurs through the desorption of these molecules from dust ice mantles, as well as the desorption of precursor molecules that form in the gas phase and subsequently react to form these species. 
This approach allows for the continued formation of carbon-bearing molecules, even in the later stages when carbon atoms and ions have formed stable CO, which would otherwise inhibit the formation of such molecules.
     \item We investigated the influence of various C/O values on molecule formation within TMC-1 based on our model and mechanisms. Previous studies have shown that different C/O ratios affect the total number of species whose simulated abundances agree with observations (the fraction of reproduced species). However, our simulation results indicate that varying the C/O values has almost no impact on the total number of molecules that fit the observations, although it does correspond to different evolutionary timescales for the molecular cloud (as show in Figure \ref{fig:Fractionofpercc/o}). This is mainly because the introduction of new mechanisms has altered the distribution of species between the dust and gas phases, as discussed in conclusions 3.
    \end{enumerate}

\begin{acknowledgments}
This work was supported by the National Natural Science Foundation of China (Grant No. 12373026), the Leading Innovation and Entrepreneurship Team of Zhejiang Province of China (Grant No. 2023R01008),  the Key R\&D Program of Zhejiang, China (Grant No. 2024SSYS0012) and the China Postdoctoral Science Foundation under grant No. 2023M733271.

\end{acknowledgments}

\appendix
\section{appendix A}

\label{appendix A}
Table \ref{Suprathermalreaction} shows the two-body suprathermal reactions occurring on both the dust surface and within the ice layer, as utilized in this paper. 
The original two-body reaction network for dust, presented in Table \ref{Suprathermalreaction}, is derived from our previous work \cite{2018ApJ...869..165L}.
This reaction network is based on the work of \cite{2011AA...530A..61H}, with additional reactions incorporated, as detailed in Section 2.2.2 of \cite{2018ApJ...869..165L}.
This compilation is derived from two-body reactions on the dust surface and the species capable of being excited (as detailed in Table 3). In total, there are 343 suprathermal reactions on the dust.
\startlongtable
\clearpage
\begin{deluxetable*}{l}
\tablecaption{Two-Body Excited Reactions on Dust}
\tabletypesize{\scriptsize}
\tablehead{
\colhead{Suprathermal Reactions} 
} 
\label{Suprathermalreaction}
\startdata
JO$^*$ (28 reactions) 		\\
\cline{1-1}
JO$^*$                   +     JC                   $\rightarrow$ JCO              \\
JO$^*$                   +     JH                   $\rightarrow$ JOH                \\  
JO$^*$                   +     JN                   $\rightarrow$ JNO                  \\
JO$^*$                   +     JC$_2$           $\rightarrow$ JCCO                 \\
JO$^*$                   +     JC$_3$            $\rightarrow$ JC$_3$O                 \\
JO$^*$                   +     JCH                  $\rightarrow$ JHCO                 \\
JO$^*$                   +     JCH$_2$          $\rightarrow$ JH$_2$CO             \\
JO$^*$                   +     JCH$_3$         $\rightarrow$ JCH$_2$OH             \\
JO$^*$                   +     JCN                  $\rightarrow$ JOCN               \\
JO$^*$                   +     JCO                  $\rightarrow$ JCO$_2$                 \\
JO$^*$                   +     JCS                  $\rightarrow$ JOCS                 \\
JO$^*$                   +     JHCO                 $\rightarrow$ JCO$_2$                 +     JH          \\      
JO$^*$                   +     JHCO                 $\rightarrow$ JCO                  +     JOH          \\     
JO$^*$                   +     JHNO                 $\rightarrow$ JNO                  +     JOH            \\     
JO$^*$                   +     JHS                  $\rightarrow$ JSO                  +     JH                  \\
JO$^*$                   +     JNH                  $\rightarrow$ JHNO	\\
JO$^*$                   +     JNH$_2$          $\rightarrow$ JHNO                 +     JH        \\          
JO$^*$                   +     JNS                  $\rightarrow$ JNO                  +     JS            \\      
JO$^*$                   +     JO                   $\rightarrow$ JO$_2$	\\
JO$^*$                   +     JO$_2$            $\rightarrow$ JO$_3$	\\	
JO$^*$                   +     JO$_2$H           $\rightarrow$ JO$_2$                  +     JOH               \\
JO$^*$                   +     JOH                  $\rightarrow$ JO$_2$H                    \\    
JO$^*$                   +     JS                   $\rightarrow$ JSO                        \\ 
JO$^*$                   +     JSO                  $\rightarrow$ JSO$_2$                      \\  
JO$^*$                   +     JCH$_4$            $\rightarrow$ JCH$_3$OH                       \\
JO$^*$                   +     JCH$_4$             $\rightarrow$ JH$_2$CO                +     JH$_2$         \\   
JO$^*$                   +     JCH$_3$OH       $\rightarrow$ JCH$_3$                 +     JHCO               \\
JO$^*$                   +     JNO                  $\rightarrow$ JNO$_2$\\
\cline{1-1}
JH$_2^*$ (22 reactions) \\
\cline{1-1}
JH$_2^*$                  +     JC                   	$\rightarrow$ JCH$_2$  \\
JH$_2^*$                  +     JC$_2$                  $\rightarrow$ JC$_2$H                 +     JH\\
JH$_2^*$                  +     JC$_2$H                 $\rightarrow$ JC$_2$H$_2$             +     JH\\
JH$_2^*$                  +     JC$_3$                  $\rightarrow$ JC$_3$H                 +     JH\\
JH$_2^*$                  +     JC$_3$H                 $\rightarrow$ JC$_3$H$_2$             +     JH\\
JH$_2^*$                  +     JC$_4$                  $\rightarrow$ JC$_4$H                 +     JH\\
JH$_2^*$                  +     JC$_4$H                 $\rightarrow$ JC$_4$H$_2$             +     JH\\
JH$_2^*$                  +     JC$_5$                  $\rightarrow$ JC$_5$H                 +     JH\\
JH$_2^*$                  +     JC$_5$H                 $\rightarrow$ JC$_5$H$_2$             +     JH\\
JH$_2^*$                  +     JC$_6$                  $\rightarrow$ JC$_6$H                 +     JH\\
JH$_2^*$                  +     JC$_6$H                 $\rightarrow$ JC$_6$H$_2$             +     JH\\
JH$_2^*$                  +     JC$_7$                  $\rightarrow$ JC$_7$H                 +     JH\\
JH$_2^*$                  +     JC$_7$H                 $\rightarrow$ JC$_7$H$_2$             +     JH\\
JH$_2^*$                  +     JC$_8$                  $\rightarrow$ JC$_8$H                 +     JH\\
JH$_2^*$                  +     JC$_8$H                 $\rightarrow$ JC$_8$H$_2$              +     JH\\
JH$_2^*$                  +     JC$_9$                  $\rightarrow$ JC$_9$H                 +     JH\\
JH$_2^*$                  +     JC$_9$H                 $\rightarrow$ JC$_9$H$_2$             +     JH\\
JH$_2^*$                  +     JCH$_2$                 $\rightarrow$ JCH$_3$                 +     JH\\
JH$_2^*$                  +     JCH$_3$                 $\rightarrow$ JCH$_4$                 +     JH\\
JH$_2^*$                  +     JCN                  	$\rightarrow$ JHCN                    +     JH\\
JH$_2^*$                  +     JNH$_2$                 $\rightarrow$ JNH$_3$                 +     JH\\
JH$_2^*$                  +     JOH                  	$\rightarrow$ JH$_2$O                 +     JH\\
\cline{1-1}
JOH$^*$ (11 reactions)\\
\cline{1-1}
JOH$^*$                  +     JC                   $\rightarrow$ JCO                  +     JH \\
JOH$^*$                  +     JH                   $\rightarrow$ JH$_2$O                 \\
JOH$^*$                  +     JH$_2$               $\rightarrow$ JH$_2$O              +     JH \\
JOH$^*$                  +     JO                   $\rightarrow$ JO$_2$H                 \\
JOH$^*$                  +     JCH$_2$              $\rightarrow$ JCH$_2$OH               \\
JOH$^*$                  +     JCH$_3$              $\rightarrow$ JCH$_3$OH               \\
JOH$^*$                  +     JCO                  $\rightarrow$ JCO$_2$              +     JH\\
JOH$^*$                  +     JH$_2$CO             $\rightarrow$ JHCO                 +     JH$_2$O\\
JOH$^*$                  +     JHCO                 $\rightarrow$ JCH$_2$O$_2$
% ...
\\
\cline{1-1}
JNH$_2^*$ (11 reactions)\\
\cline{1-1}
JNH$_2^*$                 +     JC                   $\rightarrow$ JHNC                 +     JH    \\
JNH$_2^*$                 +     JCH                  $\rightarrow$ JCH$_3$N    \\
JNH$_2^*$                 +     JCH$_2$              $\rightarrow$ JCH$_2$NH$_2$ \\
JNH$_2^*$                 +     JH                   $\rightarrow$ JNH$_3$\\
JNH$_2^*$                 +     JH$_2$               $\rightarrow$ JNH$_3$                 +     JH  \\
JNH$_2^*$                 +     JN                   $\rightarrow$ JN$_2$H$_2$ \\
JNH$_2^*$                 +     JCH$_3$              $\rightarrow$ JCH$_5$N\\
JNH$_2^*$                 +     JHCO                 $\rightarrow$ JNH$_2$CHO\\
JNH$_2^*$                 +     JNO                  $\rightarrow$ JH$_2$O                 +     JN$_2$\\
JNH$_2^*$                 +     JO                   $\rightarrow$ JHNO                 +     JH\\
JNH$_2^*$                 +     JOH                  $\rightarrow$ JNH$_2$OH\\
\cline{1-1}
JCH$_2^*$ (16 reactions)\\
\cline{1-1}
JCH$_2^*$                 +     JC                   $\rightarrow$ JC$_2$H$_2$\\
JCH$_2^*$                 +     JCH                  $\rightarrow$ JC$_2$H$_3$\\
JCH$_2^*$                 +     JCH$_2$                 $\rightarrow$ JC$_2$H$_4$\\
JCH$_2^*$                 +     JCH$_3$                 $\rightarrow$ JC$_2$H$_5$\\
JCH$_2^*$                 +     JCN                  $\rightarrow$ JC$_2$H$_2$N\\
JCH$_2^*$                 +     JHNO                 $\rightarrow$ JCH$_3$                 +     JNO\\
JCH$_2^*$                 +     JNH$_2$                 $\rightarrow$ JCH$_2$NH$_2$\\
JCH$_2^*$                 +     JO$_2$                  $\rightarrow$ JH$_2$CO                +     JO\\
JCH$_2^*$                 +     JH                   $\rightarrow$ JCH$_3$\\
JCH$_2^*$                 +     JH$_2$                  $\rightarrow$ JCH$_3$                 +     JH\\
JCH$_2^*$                 +     JN                   $\rightarrow$ JH$_2$CN\\
JCH$_2^*$                 +     JNH                  $\rightarrow$ JCH$_3$N\\
JCH$_2^*$                 +     JO                   $\rightarrow$ JH$_2$CO\\
JCH$_2^*$                 +     JOH                  $\rightarrow$ JCH$_2$OH\\
JCH$_2^*$                 +     JCH$_3$OH               $\rightarrow$ JCH$_3$OCH$_3$\\
JCH$_2^*$                 +     JCH$_3$OH               $\rightarrow$ JC$_2$H$_5$OH\\
\cline{1-1}
JCH$_3^*$ (20 reactions)\\
\cline{1-1}
JCH$_3^*$                 +     JC                   $\rightarrow$ JC$_2$H$_3$\\
JCH$_3^*$                 +     JCH                  $\rightarrow$ JC$_2$H$_4$\\
JCH$_3^*$                 +     JCH$_2$              $\rightarrow$ JC$_2$H$_5$\\
JCH$_3^*$                 +     JC$_3$N              $\rightarrow$ JCH$_3$C$_3$N\\
JCH$_3^*$                 +     JC$_5$N              $\rightarrow$ JCH$_3$C$_5$N\\
JCH$_3^*$                 +     JC$_7$N              $\rightarrow$ JCH$_3$C$_7$N\\
JCH$_3^*$                 +     JCH$_2$OH            $\rightarrow$ JC$_2$H$_5$OH\\
JCH$_3^*$                 +     JCH$_3$O            $\rightarrow$ JCH$_3$OCH$_3$\\
JCH$_3^*$                 +     JCH$_3$              $\rightarrow$ JC$_2$H$_6$\\
JCH$_3^*$                 +     JCN                  $\rightarrow$ JC$_2$H$_3$N\\
JCH$_3^*$                 +     JHCO                 $\rightarrow$ JCH$_3$CHO\\
JCH$_3^*$                 +     JHNO                 $\rightarrow$ JCH$_4$                 +     JNO\\
JCH$_3^*$                 +     JH                   $\rightarrow$ JCH$_4$\\
JCH$_3^*$                 +     JH$_2$               $\rightarrow$ JCH$_4$                 +     JH\\
JCH$_3^*$                 +     JN                   $\rightarrow$ JCH$_3$N\\
JCH$_3^*$                 +     JNH                  $\rightarrow$ JCH$_3$NH\\
JCH$_3^*$                 +     JNH$_2$              $\rightarrow$ JCH$_5$N\\
JCH$_3^*$                 +     JO                   $\rightarrow$ JCH$_2$OH\\
JCH$_3^*$                 +     JOH                  $\rightarrow$ JCH$_3$OH\\
JCH$_3^*$                 +     JS                   $\rightarrow$ JH$_2$CS                +     JH\\
\cline{1-1}
JCH$_4^*$ (4 reactions)\\
\cline{1-1}
JCH$_4^*$                 +     JC$_2$H                 $\rightarrow$ JC$_2$H$_2$                +     JCH$_3$\\
JCH$_4^*$                 +     JH                   $\rightarrow$ JCH$_3$                 +     JH$_2$\\
JCH$_4^*$                 +     JO                   $\rightarrow$ JCH$_3$OH\\
JCH$_4^*$                 +     JO                   $\rightarrow$ JH$_2$CO                +     JH$_2$\\
\cline{1-1}
JHCO$^*$ (8 reactions)\\
\cline{1-1}
JHCO$^*$                 +     JCH$_3$              $\rightarrow$ JCH$_3$CHO\\
JHCO$^*$                 +     JH                   $\rightarrow$ JH$_2$CO\\
JHCO$^*$                 +     JNH$_2$              $\rightarrow$ JNH$_2$CHO\\
JHCO$^*$                 +     JO                   $\rightarrow$ JCO$_2$              +     JH\\
JHCO$^*$                 +     JO                   $\rightarrow$ JCO                  +     JOH\\
JHCO$^*$                 +     JOH                  $\rightarrow$ JCH$_2$O$_2$\\
JHCO$^*$                 +     JCH$_2$OH            $\rightarrow$ JCH$_2$OHCHO\\
JHCO$^*$                 +     JCH$_3$O             $\rightarrow$ JHCOOCH$_3$\\
\cline{1-1}
JH$_2$CO$^*$ (3 reactions)\\
\cline{1-1}
JH$_2$CO$^*$                +     JH                   $\rightarrow$ JHCO                 +     JH$_2$\\
JH$_2$CO$^*$                +     JH                   $\rightarrow$ JCH$_2$OH\\
JH$_2$CO$^*$                +     JOH                  $\rightarrow$ JHCO                 +     JH$_2$O\\
\cline{1-1}
JCH$_3$O$^*$ (4 reactions)\\
\cline{1-1}
JCH$_3$O$^*$                +     JHCO                 $\rightarrow$ JHCOOCH$_3$\\
JCH$_3$O$^*$                +     JCH$_3$                 $\rightarrow$ JCH$_3$OCH$_3$\\
JCH$_3$O$^*$                +     JH                   $\rightarrow$ JCH$_3$OH\\
JCH$_3$O$^*$                +     JN                   $\rightarrow$ JNH$_2$CHO\\
\cline{1-1}
JCH$_2$OH$^*$ (5 reactions)\\
\cline{1-1}
JCH$_2$OH$^*$               +     JCH$_3$                 $\rightarrow$ JC$_2$H$_5$OH\\
JCH$_2$OH$^*$               +     JCH$_3$                 $\rightarrow$ JCH$_3$OCH$_3$\\
JCH$_2$OH$^*$               +     JH                   $\rightarrow$ JCH$_3$OH\\
JCH$_2$OH$^*$               +     JN                   $\rightarrow$ JNH$_2$CHO\\
JCH$_2$OH$^*$               +     JHCO                 $\rightarrow$ JCH$_2$OHCHO\\
\cline{1-1}
JCH$_3$OH$^*$ (5 reactions)\\
\cline{1-1}
JCH$_3$OH$^*$               +     JC                   $\rightarrow$ JCH$_3$CHO   \\
JCH$_3$OH$^*$               +     JC                   $\rightarrow$ JCH$_3$                 +     JHCO \\
JCH$_3$OH$^*$               +     JO                   $\rightarrow$ JCH$_3$                 +     JHCO  \\
JCH$_3$OH$^*$               +     JCH$_2$                 $\rightarrow$ JCH$_3$OCH$_3$ \\
JCH$_3$OH$^*$               +     JCH$_2$                 $\rightarrow$ JC$_2$H$_5$OH\\
\enddata
\end{deluxetable*}

\begin{figure*}
	\plotone{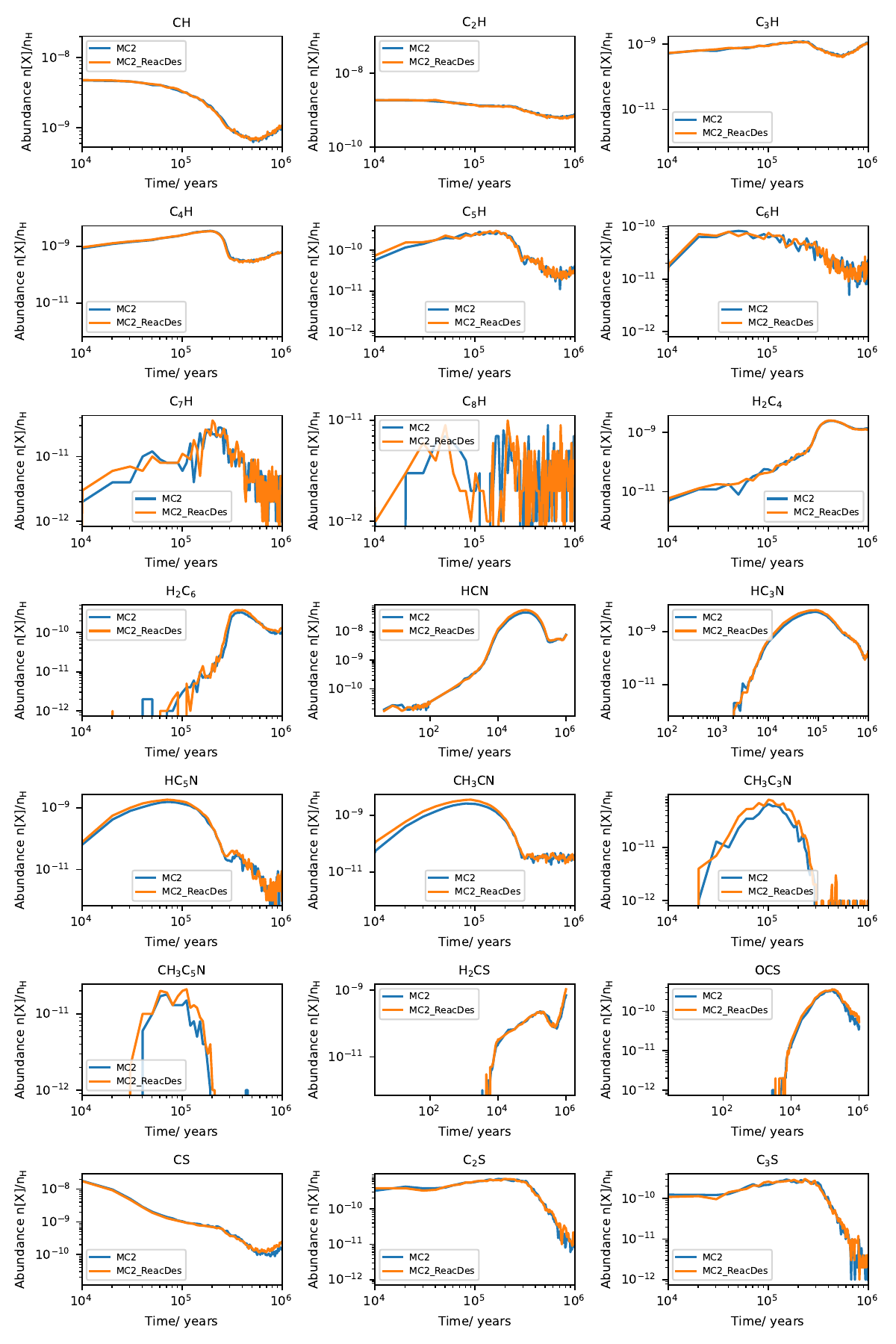}
    \caption{Abundance Ratios of Gas Carbon Chain Species Relative to H in MC2 model and MC2\_reacDes model.}
    \label{fig:carbonone_reades}
\end{figure*}

\begin{figure*}
	\plotone{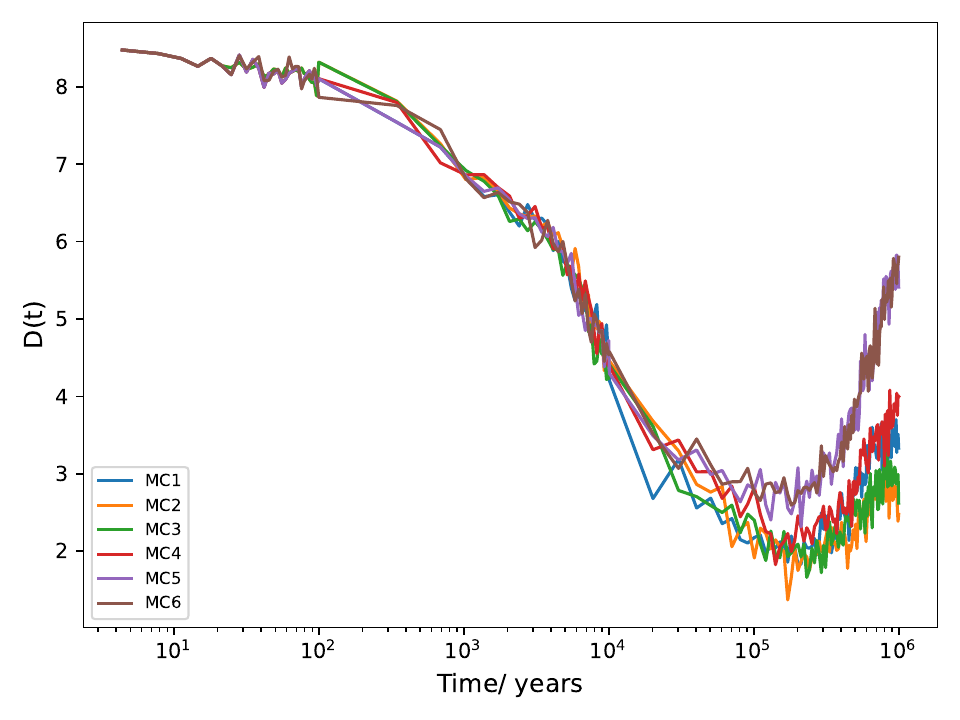}
    \caption{Variation of D(t) over time across different models.}
    \label{fig:Dt}
\end{figure*}

In rate equation models, such as those used in Nautilus \citep{2016MNRAS.459.3756R}, species abundances are assigned a lower limit of $1 \times 10^{-40}$. Similarly, we used equation \ref{eq_best_local} to calculate the global best-fitting time, setting a lower limit of $1 \times 10^{-40}$ 
for molecular abundances of 0. Figure \ref{fig:Dt} shows the relationship between D(t) and time across different models. The minimum D(t) values for the MC1 to MC6 models are $2.1 \times 10^5$,  $1.7 \times 10^5$,  $2.3 \times 10^5$,  $1.4 \times 10^5$, 
 $2.1 \times 10^5$, and  $1.8 \times 10^5$ years, respectively.
 
\bibliography{sample631}{}
\bibliographystyle{aasjournal}

\end{document}